\documentclass[fleqn,usenatbib,useAMS]{mnras}

\usepackage{newtxtext,newtxmath}
 
\usepackage[T1]{fontenc}
\usepackage{graphicx}	
\usepackage{amsmath}	
\usepackage{amssymb}	
\usepackage{bm}		
\usepackage{float}

\title[SMGs in two massive galaxy protoclusters at $z=2.24$]{Submillimetre galaxies in two massive protoclusters at $z=2.24$:  witnessing the enrichment of extreme starbursts in the outskirts of HAE density peaks}
\author[Y.~H.~Zhang et al.]
{Yuheng~Zhang,$^{1,2}$
Xian~Zhong~Zheng,$^{1,2}$\thanks{E-mail: xzzheng@pmo.ac.cn}
Dong~Dong~Shi,$^{1}$
Yu~Gao,$^{3,1}$
Helmut Dannerbauer,$^{4,5}$
Fang~Xia~An,$^{6}$
\and
Xinwen~Shu,$^{7}$
Zhen-Kai~Gao,$^{8}$
Wei-Hao~Wang,$^{8}$
Xin~Wang,$^{9}$
Zheng~Cai,$^{10}$
Xiaohui~Fan,$^{11}$
Min~Fang,$^{1,2}$
\and
Zhizheng~Pan,$^{1,2}$
Wenhao~Liu,$^{1,2}$
Qinghua~Tan,$^{1,2}$
Jianbo~Qin,$^{1}$
Jian~Ren,$^{1,2}$
Man~Qiao,$^{1,2}$
Run~Wen$^{1,2}$
\and
and Shuang~Liu$^{1,2}$
\\
$^{1}$Purple Mountain Observatory, Chinese Academy of Sciences, 10 Yuanhua Road, Nanjing 210023, People's Republic of China\\
$^{2}$School of Astronomy and Space Sciences, University of Science and Technology of China, Hefei 230026, People's Republic of China\\
$^{3}$Department of Astronomy, Xiamen University, 1 Zengcuoan West Road, Xiamen, Fujian 361005, People's Republic of China\\
$^{4}$Instituto de Astrofísica de Canarias (IAC), E-38205 La Laguna, Tenerife, Spain\\
$^{5}$Universidad de La Laguna, Dpto. Astrofísica, E-38206 La Laguna, Tenerife, Spain\\
$^{6}$Inter-University Institute for Data Intensive Astronomy, and Department of Physics and Astronomy, University of the Western Cape, \\ \ \ Robert Sobukwe Road, Bellville, Cape Town 7535, South Africa \\
$^{7}$Department of Physics, Anhui Normal University, Wuhu, Anhui 241000, People's Republic of China\\
$^{8}$Academia Sinica Institute of Astronomy and Astrophysics (ASIAA), No. 1, Section 4, Roosevelt Rd., Taipei 10617, Republic of China\\
$^{9}$Infrared Processing and Analysis Center, Caltech, 1200 E. California Blvd., Pasadena, CA 91125, USA\\
$^{10}$Department of Astronomy, Tsinghua University, Beijing 100084, People's Republic of China\\
$^{11}$Steward Observatory, University of Arizona, 933 North Cherry Avenue, Tucson, AZ 85721, USA
}

\date{Accepted 2022 March 19.  Received 2022 February 25;  in original form 2021 November 26}

\begin{document}
\label{firstpage} 
\pagerange{\pageref{firstpage}--\pageref{lastpage}}
\maketitle
\begin{abstract}

Submillimetre galaxies represent a rapid growth phase of both star formation and massive galaxies. Mapping SMGs in galaxy protoclusters  provides key insights into where and how these extreme starbursts take place in connections with the assembly of the large-scale structure in the early Universe. We search for SMGs at 850\,$\micron$ using JCMT/SCUBA-2 in two massive protoclusters at $z=2.24$, BOSS1244 and BOSS1542, and detect 43 and 54 sources with $S_{850}>4$\,mJy at the $4\sigma$ level within an effective area of 264\,arcmin$^2$, respectively.  We construct the intrinsic number counts and find that the abundance of SMGs is $2.0\pm0.3$ and $2.1\pm0.2$ times that of the general fields, 
confirming that BOSS1244 and BOSS1542 contain a higher fraction of dusty galaxies with strongly enhanced star formation. The volume densities of the SMGs are estimated to be $\sim15-$30 times the average, significantly higher than the overdensity factor ($\sim 6$) traced by H$\alpha$ emission-line galaxies (HAEs). More importantly,  we discover a prominent offset between the spatial distributions of the two populations in these two protoclusters --- SMGs are mostly located around the high-density regions of HAEs, and few are seen inside these regions. This finding may have revealed for the first time the occurrence of  violent star formation enhancement  in the outskirts of the HAE density peaks, likely driven by the boosting of gas supplies and/or starburst triggering events. Meanwhile, the lack of SMGs inside the most overdense regions  at $z\sim2$ implies a transition to the environment disfavouring extreme starbursts.

\end{abstract}





\section{INTRODUCTION}
\label{1:intro}

Characterizing galaxy properties in different environments across cosmic time is essential to building up a complete picture of galaxy formation and evolution, and understanding when, where and how the driving physical processes take place \citep{Somerville2015}. A large portion of  stars in the local universe were formed at the peak epoch ($z\sim2-$3) of cosmic star formation and active galactic nucleus (AGN) activities \citep[e.g.][]{Hopkins2006, Zheng2009, Madau2014}. Galaxy protoclusters at $z>2$, as the progenitors of local galaxy clusters formed at the densest nodes of the cosmic web,  provide a direct probe of the rapid assembly of cosmic structures, and insights into the formation of massive galaxies in relation to the structure formation at this epoch \citep[e.g.][]{DeLucia2007}. Mapping  galaxy populations from extreme starbursts to quiescent ones in these protoclusters enables us to dissect the complex processes regulating galaxy evolution and the influence of the local and global environments,  making the $z\sim2-$3 protoclusters unique testbeds for the theoretical models of galaxy formation and evolution \citep{Overzier2016}.

Submillimetre galaxies (SMGs) are ultraluminous dusty star-forming galaxies (SFGs) with the vast majority of radiation energy in the far-infrared (FIR) and sub-millimetre bands \citep[][]{Smail1997, Barger1998, Hughes1998, Michalowski2012, Casey2014}.  They are extreme starbursts with high star formation rate (SFR) of $\sim10^2$$-$$10^3$\,M$_{\odot}$\,yr$^{-1}$ and high stellar mass of $\sim10^{10-11}$\,M$_\odot$,  located preferentially at $z\sim 2-$3 \citep[][and references therein]{Chapman2005, Wardlow2011, Smolcic2012, Simpson2014, Smith2017, Michalowski2017, Danielson2017, Brisbin2017,Hodge2020}.  The physical properties of SMGs derived from detailed studies of individual sources and large sky area submillimetre surveys suggest that SMGs represent an early evolutionary phase of all local  ellipticals 
 \citep{Smail2002, Swinbank2006, Fu2013, Toft2014, Ikarashi2015, Miettinen2017, An2019, Gullberg2019, Dudzeviciute2020, Rennehan2020}. Although SMGs are a rare population \citep[$\sim$400\,deg$^{-2}$ down to $S_{850}=4$\,mJy;][]{Simpson2019, Shim2020}, they contribute a significant fraction ($\sim$20\,per\,cent) of the cosmic SFR density at $z>2$ \citep{Bourne2017,Koprowski2017, Zavala2021}. Their extreme SFRs are correlated with higher gas fractions compared to normal SFGs at the same epoch \citep{Bothwell2013, Scoville2016, Decarli2016, Tacconi2018}.  The large amount gas is thought to be supplied by gas infall via cold streams from surrounding gas reservoirs \citep{Narayanan2015, Ginolfi2017}. The extreme starbursts of this population are partially triggered by galaxy major mergers \citep[][but see \citealt{Dave2010, Narayanan2015, McAlpine2019}]{Tacconi2008, Engel2010}, while a diversity of morphologies unveiled from the rest-frame ultraviolet (UV) and optical imaging indicate galaxy interactions and disk instabilities to be important mechanisms for enhancing star formation in SMGs  \citep{Swinbank2010, Kartaltepe2012, Chen2015}, as well as AGN activities \citep[][]{Chapman2005,Wang2013}.  Because the rarity and  high SFR of SMGs are sensitive to the physical processes governing galaxy formation (e.g., star formation, stellar and AGN feedback, gas infall, metal enrichment and galaxy merging/interactions),  the SMG population is used to constrain cosmological models. It remains challenging to reproduce the SMG population with high SFRs matching observations at high $z$ \citep{Dave2010, McAlpine2019, Hayward2021, Lovell2021}. 

 Moreover, SMGs are found to reside in massive haloes including high-$z$ radio galaxies (HzRGs) \citep{Stevens2003, Humphrey2011, Rigby2014, Dannerbauer2014, Zeballos2018}, quasars  \citep{Jones2017, Wethers2020} and enormous Ly$\alpha$ nebulae (ELANs) \citep{Arrigoni2018, Nowotka2022} in terms of their clustering \citep{Blain2004, Hickox2012, Wilkinson2017}. 
These  massive haloes trace the dense large-scale structures (and preferentially galaxy protoclusters) in the high-$z$ universe.  Indeed, SMGs were often detected in  high-$z$ protoclusters of galaxies, closely linked with the early formation of massive galaxies in the cluster core regions \citep{Venemans2007, Chapman2009, Galametz2013, Dannerbauer2014, Umehata2015, Coogan2018, Lacaille2019}.  However, the abundance and birthplace of SMGs rely on the evolutionary stage of  galaxy protoclusters and how massive galaxies form in and around the protoclusters \citep{Shimakawa2018, Cooke2019, Shi2020}. On the other hand, large-scale environment is reported to play a key role in setting of the circumgalactic medium (CGM) angular momentum that governs gas infall and star formation in a galaxy \citep{Wang2021, Lu2021}.  Therefore, mapping SMGs in and around the high-$z$ protoclusters of galaxies is strongly demanded to uncover where and how extreme starbursts take place in the densest large-scale structures, and shed light on how massive galaxies form in the overdense environments \citep{Bahe2017, Bassini2020, Lim2021}.

Galaxy protoclusters at $z>2$ are usually unviralized and extended over a scale of $\sim 20\,h^{-1}$\,co-moving\,Mpc \citep[cMpc;][]{Chiang2013, Muldrew2015, Lovell2018}.  
Using a novel approach based on groups of coherently strong Ly$\alpha$ absorption (CoSLA) imprinted on the spectra of a number of background quasars,  i.e.,  named the Mapping the Most Massive Overdensities Through Hydrogen \citep[MAMMOTH;][]{Cai2016, Cai2019}, several extremely massive overdensity regions at $z=2-$3 of scales of $\sim30\,h^{-1}$\,cMpc have been spectroscopically confirmed, including BOSS1441 at $z=2.32$ \citep{Cai2017}, BOSS1244 and BOSS1542 at $z=2.24$ \citep{Zheng2021, Shi2021}. The latter two MAMMOTH overdensities are found to exhibit distinct morphologies --- BOSS1244 consists of three components in assembly and BOSS1542 is a first giant filamentary structure at $z>2$ discovered to date --- suggestive of being in different assembling stages.

In this paper, we carry out  a deep 850\,$\micron$ survey of  BOSS1244 and BOSS1542  with the Submillimetre Common-User Bolometer Array 2 (SCUBA-2) at the East Asian Observatory's James Clerk Maxwell Telescope (JCMT) to pinpoint SMGs in the two most massive overdensities at $z=2.24$,  and explore their properties and connections with  other galaxy populations.  In Section~\ref{sec:data}, We briefly describe our observations and data reduction. We  present the procedures for source extraction along with the completeness estimate, flux boosting factor, position offset and the false-detection rate with Monte Carlo simulations. In Section~\ref{sec:results}, we present our results on number counts,  the overdensity estimate, and the spatial distribution of detected SMGs. Discussion and summary of our results are given in Section~\ref{sec:summary}. A standard $\Lambda$CDM cosmology with $H_0=70$\,km$^{-1}$\,Mpc$^{-1}$, $\Omega _{\rm \Lambda}$=0.7 is adopted throughout the paper, giving a scale of 0.495\,physical\,Mpc (pMpc) or 1.602\,comoving\,Mpc (cMpc)\,per\,arcmin at $z = 2.24$.

\begin{figure*}
    \includegraphics[width=\columnwidth]{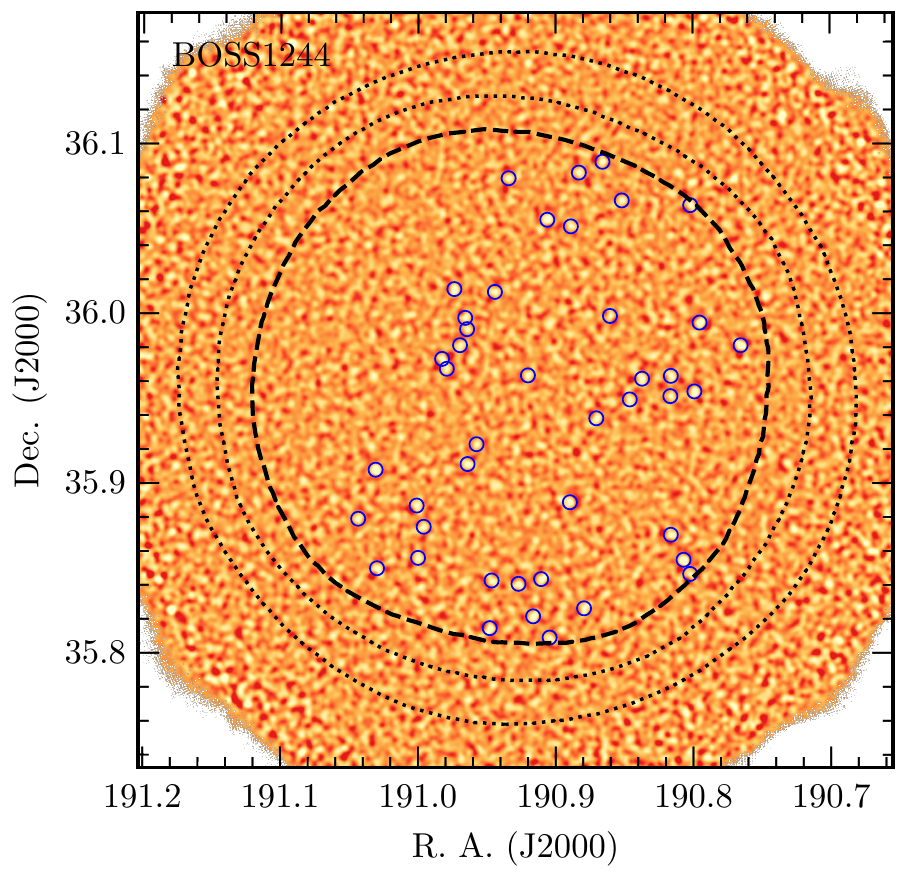}
    \includegraphics[width=\columnwidth]{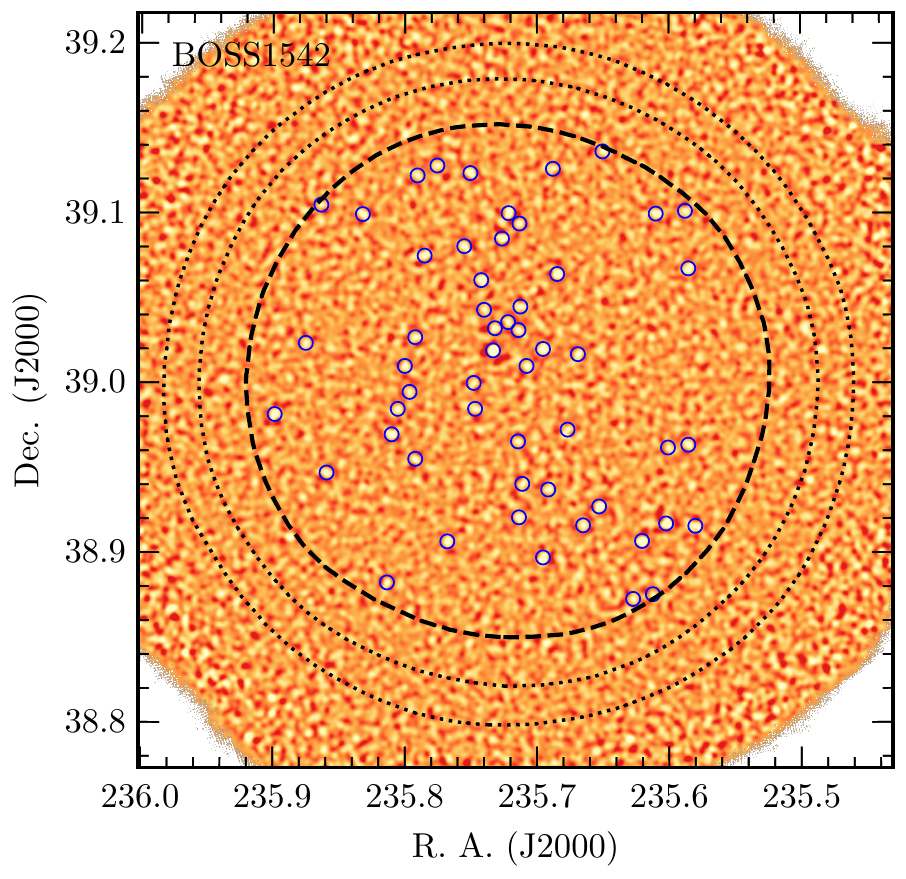}
    \caption{Deep SCUBA-2 850\,$\micron$ mosaic maps of two MAMMOTH fields BOSS1244 (left) and BOSS1542 (right). Blue circles mark 43 and 54 detections  at the level of S/N$\geq$4 ($S_{850}\geq 4$\,mJy) within the effective area enclosed by the thick dashed lines. The dotted lines enclose the coverage where the noise levels are 2 and 3\,mJy from inside to out. There are 35 and 39 sources detected with $3.5\leq {\rm S/N}<4$ in the effective area of BOSS1244 and BOSS1542, respectively. Due to a rapidly increasing rate of false sources at S/N $<4$, we take sources with S/N$\geq$4 as our SMG samples.}
    \label{fig:snr}
\end{figure*}

\section{DATA REDUCTIONS AND ANALYSIS} \label{sec:data}

\subsection{Observations} \label{sec:observations}

Observations of our deep 850\,$\micron$ survey of two target fields BOSS1244 and BOSS1542 were performed in February and June of 2018,  March and April of  2019 (program IDs: M18AP025 and M19AP028) with the SCUBA-2 instrument on board JCMT \citep{Holland2013}. About half of the observations were done under the band-3 weather condition ($0.05<\tau _{\rm 225GHz} \leq 0.08$), while the rest were taken under the band-2  weather condition ($0.08<\tau _{\rm 225GHz} \leq 0.012$). The average optical depth at  850\,$\micron$ was approximately $\tau_{850\micron} \approx 0.34$.

The SCUBA-2  observing mode of PONG-900 was adopted to scan the density peak area in BOSS1244 (centred at $\alpha=12:43:55.49$, $\delta=+35:59:37.4$)  and  BOSS1542 ($\alpha=15:42:19.24$, $\delta=+38:54:14.1$), respectively.  The telescope scan speed was $280\arcsec$\,s$^{-1}$ with 11 rotations during an  integration time of 40 minutes to ensure a uniform sensitivity map within a circle of a $15\arcmin$ diameter. Finally, there are 28 and 32 useful repeated scan maps with a total effective integration time of 18.67 and 21.33\,hours in BOSS1244 and BOSS 1542, respectively.

\begin{table*} 
\centering
\caption{A summary of our SCUBA-2 observations for BOSS1244 and BOSS1542.} \label{tab:tab1}
\begin{tabular}{cccccccc}
    \hline \hline
    Field  &  R.A.  & Dec. & Scan Pattern &  Exposure &  Central Noise ($\sigma_{\rm CN}$) &  Effective Area \\
    & (J2000.0) & (J2000.0) & & (hr) & (mJy\,beam$^{-1}$) & (arcmin$^{2}$) \\
    \hline
    BOSS1244 & 12:43:55.49 & +35:59:37.4 & PONG 900  & 18.67     & 1.07      & 263.5 \\
    BOSS1542 & 15:42:19.24 & +38:54:14.1 & PONG 900  & 21.33     & 0.98      & 263.7 \\
    \hline
\end{tabular}\\
Note: Central noise refers to the square root of the minimum value in the central region of the variance map. \\
The effect area is the region where the rms sensitivity less than $1.5 \times \sigma_{\rm CN}$. \\
\end{table*}

\begin{figure*}
     \includegraphics[width=\columnwidth]{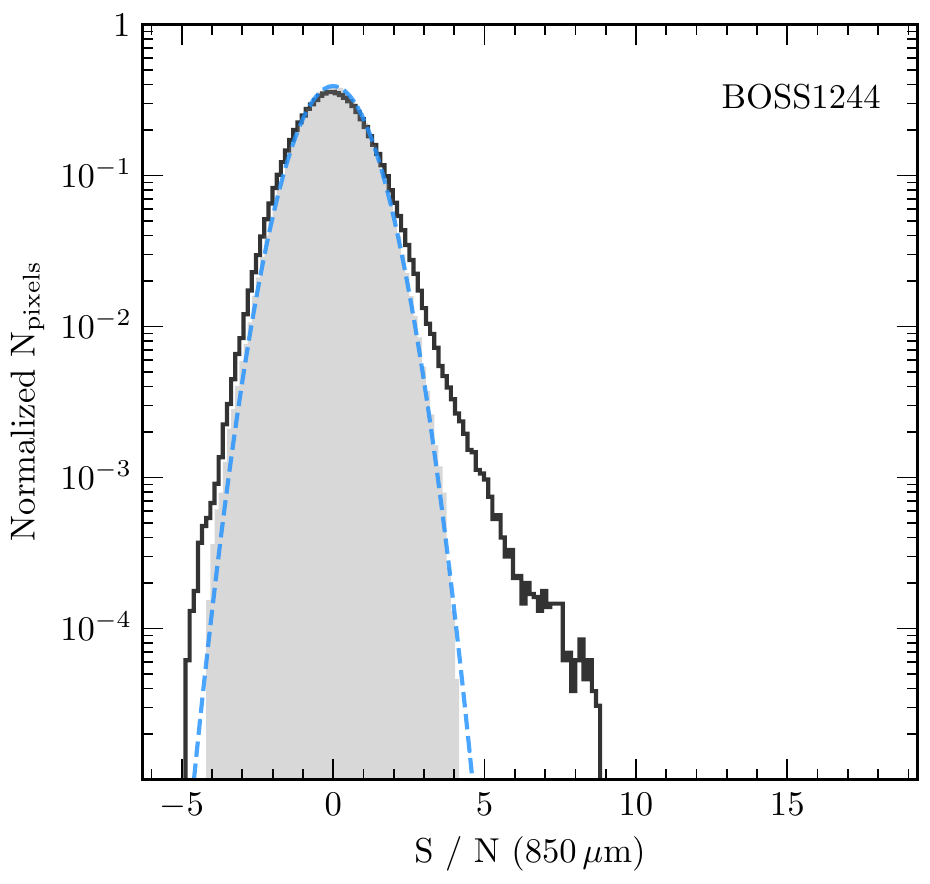}
     \includegraphics[width=\columnwidth]{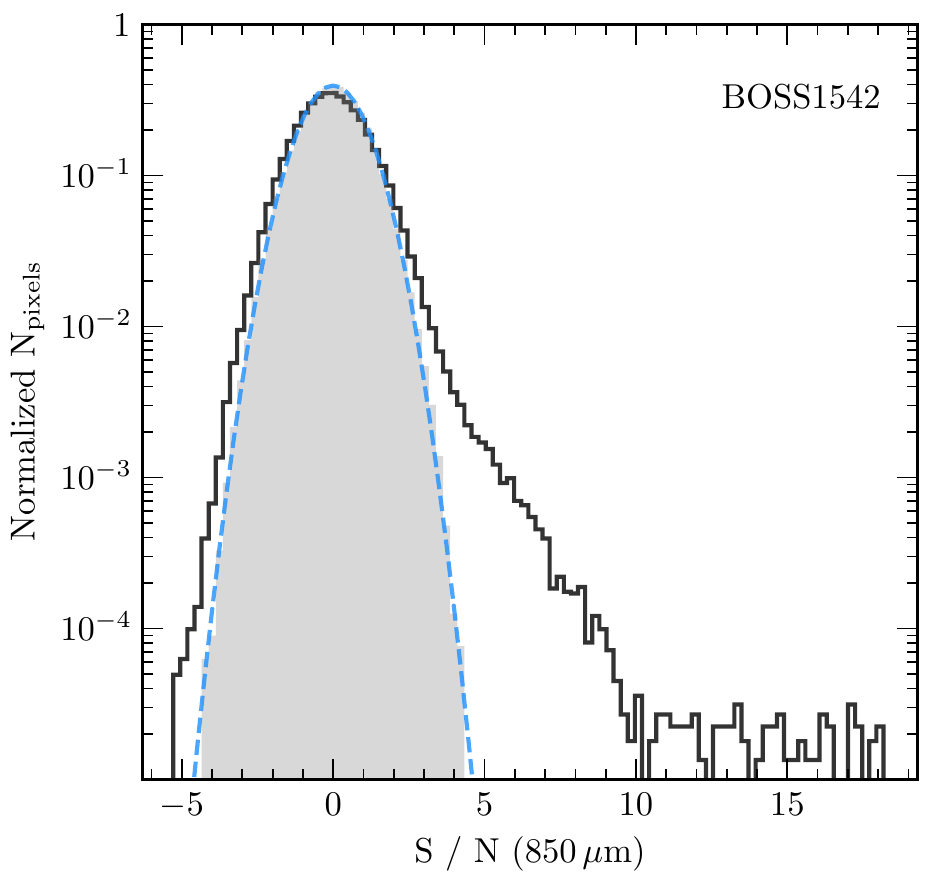}
     \caption{Histograms of pixels within the effective area in the  maps of S/N (black solid) and jackknife noise (shaded region) for BOSS1244 (left) and BOSS1542 (right). The noise distribution agrees well with a Gaussian function plotted by the blue dashed line, indicating that the instrument noise is purely random. The positive excess is attributed to astronomical signals and the negative excess is due to troughs around the bright sources introduced by matched-filtering process.}
     \label{fig:hist}
\end{figure*}

\subsection{Data reduction} \label{sec:reduction}

We use two software tools,  Sub-Millimetre User Reduction Facility (SMURF) and the Pipeline for Combining and Analysing Reduced Data (PICARD) which intergrated into the STARLINK software package, to perform our data reduction. Following  \cite{Chen2013}, we firstly used the Dynamic Iterative Map Maker (DIMM) in SMURF with the standard configuration of "$dimmconfig\_blank\_field.lis$" to reduce and stack each scan of our 850\,$\micron$ observations.  The standard reduction procedures to the raw data (e.g., down-sampling, flat-fielding and white noise estimating) were automatically processed. 

We used the updated standard Flux Conversion Factor (FCF) to calibrate fluxes for our stacked scan maps. The updated FCF was derived through the continuous analysis monitoring on the calibrators for ten years since 2011 \citep{Mairs2021}, giving 516 and 495\,Jy\,pW$^{-1}$\,beam$^{-1}$ at 850\,$\micron$ for the observations implemented before and after June 30, 2018.  The extinction correction are calculated as following
\begin{equation}
    C_{\rm ext} = \exp(-\tau_{\rm 345GHz} \times {AM}), 
\end{equation}
where $AM$ refers to the airmass of given observations recorded in image headers and  $\tau_{\rm 345GHz}$ represents  the atmospheric opacity at 850\,$\micron$. We note that  $\tau_{\rm 345GHz}$ is derived from the observed atmospheric opacity at 225\,GHz. The update of FCF induces a change in the conversion from $\tau_{\rm 225GHz}$  to $\tau_{\rm 345GHz}$.  The original  $\tau_{\rm 345GHz}$ is given by
\begin{equation}
    \tau_{\rm 345GHz,ori} = 4.6\times (\tau_{\rm 225GHz} - 0.0043), 
\end{equation}
while the new  $\tau_{\rm 345GHz}$ is calculated using
\begin{equation}
    \tau_{\rm 345GHz, new} = 3.7 \times (\tau_{\rm 225GHz} - 0.04 + 0.2 \times \sqrt{\tau_{\rm 225GHz}}). 
\end{equation}
We recovered our stacked scan maps from the  extinction correction based on $\tau_{\rm 345GHz, ori}$, and then corrected them for the extinction with $\tau_{\rm 345GHz, new}$.  After doing so, we applied the standard FCFs to our stacked scan maps. The relative calibration accuracy of peak flux has been shown to be stable to a level of 7\,per\,cent at 850\,$\micron$ \citep{Mairs2021}.  Note that the updated FCF is comparable to the original FCF within the uncertainties and applied to the fluxes and the noises of the maps simultaneously, and will not affect the number of extracted sources  as well as the number counts. 
We applied an upward correction of 10\,per\,cent to compensate for the flux lost in the filtering \citep[see][]{Chen2013}. The same upward correction was also adopted in \cite{Geach2017}.

After the flux calibration, all scan maps of a target field were co-added to produce a mosaic science map using the software tool  PICARD  with the MOSAIC\_JCMT\_IMAGES recipe \citep{Jenness2008}. We checked the data quality of each scan map with the histogram of pixel values in the  signal-to-noise ratio (S/N) map, which is obtained by dividing the flux map by the noise map. Each scan only has a very small fraction of pixels contributed by astronomical signals as the instrument noise dominates the observed fluxes in a short-time exposure.  Therefore, the histogram obeys a Gaussian function with a little excess at the positive end. We notice that the pixel histograms of four scans in BOSS1244 taken in 2018 exhibit strong excess from a normal distribution in both sides. We suspect that poor weather conditions were responsible for this abnormal distribution of instrument noise. We discarded the four scans and co-added the rest to obtain the final mosaic science map for BOSS1244.  The removal of four scans only slightly increases the central noise  by 6\,per\,cent. 

We applied a matched filtering to mosaic science maps to enhance the strength of point sources  using the software tool PICARD with the SCUBA-2\_MATCHED\_FILTER recipe. A default Gaussian kernel with a broaden FWHM of $30\arcsec$ was used to smooth a science map.\footnote{\url{https://starlink.eao.hawaii.edu/docs/sun265.htx/sun265ss20.html}}  The smoothed map was taken as the background to be subtracted from the original one. Then the background-subtracted map was convolved with a Point Spread Function (PSF) to generate the filtered science map.  A final data cube is created to include flux map and the flux variance map over all scan maps. We obtained the S/N map by dividing the flux map by the square root of the flux variance map. The S/N map is used for source extraction subsequently.

We take the minimum value of the central region as the central noise ($\sigma_{\rm CN}$) of a flux variance map.  We take the noise level to measure the sensitivity of our observations. Our deep 850\,$\micron$ maps of BOSS1244 and BOSS1542 reach a central noise level of $1\sigma_{\rm CN}=1.07$\,mJy\,beam$^{-1}$ and 0.98\,mJy\,beam$^{-1}$, respectively. We refer the effective area as to where the instrument noise is less than $1.5 \times \sigma_{\rm CN}$. The effective area is approximately a circle with a radii of $9\farcm2$ and slight larger than the nominal coverage of $15\arcmin$ with PONG-900, corresponding to 263.5\,arcmin$^{2}$ and 263.7\,arcmin$^{2}$ for BOSS1244 and BOSS1542, respectively. The mean (median) noise levels are 1.25 (1.21)\,mJy beam$^{-1}$ and 1.15 (1.12)\,mJy beam$^{-1}$ for BOSS1244 and BOSS1542, respectively. Our source extraction is performed in the effective area of a given 850\,$\micron$ map. We list central position, scan pattern, exposure time,  central noise and effective area of our observations in Table~\ref{tab:tab1}.

\subsection{Source extraction} \label{sec:extration}

SMGs are found to have a size of $<1\arcsec$ for dust emission, corresponding to several kpc at $z>1$ \citep{Ikarashi2015, Simpson2015a, Hodge2016, Chen2017, Gullberg2019}.  
The average beam size of SCUBA-2 at 850\,$\micron$ is  $\sim 15\arcsec$. These SMGs appear as unresolved point sources in the SCUBA-2 850\,$\micron$ maps. Our  850\,$\micron$ science maps are scaled to set the peak value of a point source equal to the total flux of the source.  We firstly detect sources in the science map using the software SExtractor \citep{Bertin1996}. All detected sources with S/N$>$5 in one field are stacked to construct an empirical PSF. The empirical PSF is well modeled with the superposition of two Gaussian functions \citep{Geach2017} and have an FWHM of $\sim14\farcs9$, same as given in \cite{Simpson2019}. The PSF profiles from BOSS1244 and BOSS1542 agree with each other within 2\,per\,cent. We thus use a single empirical PSF from BOSS1542 to extract sources in the two fields.

Our source extraction is based on the commonly-used ``top-down'' algorithm, which was often adopted by SCUBA-2 deep surveys \citep[e.g.][]{Geach2017, Simpson2019, Shim2020}. \cite{Wang2017} demonstrated that the standard method of source extraction is hard to effectively identify two  sources very close to each other, and thus increase the fraction of multiply sources and overestimate the fluxes of blended sources. We find that the ``top-down'' method creates a fake source with weak flux density when it subtracts a bright source. This is because the subtraction for a bright source inversely adds some fluxes to pixels in the groove area around the bright source caused in the matched-filtering process (see Section~\ref{sec:reduction}). Following \cite{Wang2017}, we adopt an iterative procedure that acts better in resolving the mildly multiply sources to extract sources, resembling the ``CLEAN'' de-convolution in the reduction process for data obtained with radio interferometries \citep{Hogbom1974}.   

We briefly describe our iteration procedure for source extraction. Firstly, we identify a source at the peak pixel in the 850\,$\micron$ S/N map and record the coordinate and the instrument noise at the position of the peak pixel. Secondly, we subtract 5\,per\,cent of total flux of the peak-scaled empirical PSF (called CLEAN ``gain'' in literature) centred on this position as the preliminary flux of the source. The procedure is iterated until the peak S/N reaches the given detection threshold. Here we adopt a detection threshold of $\geq$3$\sigma$ to construct our preliminary catalog, which will be further refined with a higher detection threshold. In previous works, $\geq$4$\sigma$ was often adopted to decrease source contamination rate \citep[e.g.][]{Geach2017, Simpson2019}. We note that adoption of a $\geq$4$\sigma$ detection will produce a source catalogue identical to our preliminary+refined catalog. When a detection locates within the half beam (approximately $8\arcsec$) of a previously-detected source in the iteration, this detection will not be treated as a new source. The 5\,per\,cent of peak-scaled PSF is subtracted and the cleaned flux is superimposed into the existing one. The final flux of every extracted source is comprised of the cleaned fluxes and the remaining flux at its initial peak position.  
Since the iterative procedure of source extraction may also mistake the pair of sources very close to each other, this will increase the flux either and affect the boosting factor estimation (Section~\ref{sec:comp}).

Note that the PONG-900 scan mode produces a uniform depth in the inner regions of a stacked scan map, and the noise level increases rapidly with radius in the outskirts due to a rapid decline of the integrated exposure time. We adopt the median noise level of $\sigma = 1.21 (1.12)$\,mJy\,beam$^{-1}$ within the effective area in BOSS1244 and BOSS1542 (Table \ref{tab:tab1}).
We perform source extraction procedure and detect 144 sources at $>$3$\sigma$ and 43 sources at $>$4$\sigma$ in BOSS1244. In BOSS1542, we detect 161 sources at $>$3$\sigma$ and 54 sources at $>$4$\sigma$.  We demonstrate the SCUBA-2 850\,$\micron$ S/N maps of BOSS1244 and BOSS1542 along with the detected sources at $>$4$\sigma$ in Figure~\ref{fig:snr}.  Our results give an observed 850\,$\micron$ source number density of $\sim 588~(737)$\,deg$^{-2}$ down to $S_{850}=4.3~(4.0)$\,mJy at the $4\sigma$ level in BOSS1244 (BOSS1542).  We remind that the observed number density needs to be corrected for the incompleteness and flux boosting to obtain the intrinsic number density.  
For a comparison, the observed number density of 850\,$\micron$ sources in blank fields is $\sim 408~(455)$\,deg$^{-2}$ down to $S_{850}=4.3~(4.0)$\,mJy \citep{Geach2017}.
Clearly, both of our two overdensity fields contain more 850\,$\micron$ sources  than that in the blank fields over the same area. There could be the field-to-field variance due to the depth and area coverage and we will investigate the overdensity of the two protoclusters in next section. 
The excess of 850\,$\micron$ sources is likely contributed by SMGs associated with the $z=2.24$ overdensities.  We notice that our observed number densities of 850\,$\micron$ sources in BOSS1244 and BOSS1542 agree well with that in BOSS1441, where 24 sources were detected down to $S_{850}=4.5$\,mJy within an effective area of 128\,arcmin$^2$ covering the enormous Lyman-$\alpha$ nebulae MAMMOTH-1 \citep{Arrigoni2018}.

\subsection{Completeness and flux boosting} \label{sec:comp}

Next step is to derive the number counts of 850\,$\micron$ sources in our two overdensity fields in order to quantify the excess of 850\,$\micron$ sources.  A variety of observational and instrumental effects may contaminate source fluxes and profiles,  and may bias source detection in the submillimetre bands. Among them,  noises from instruments and background radiation are a dominant factor to the detection incompleteness for sources with low S/N.   
We use Monte Carlo simulations to estimate the completeness of source detection in our observations. This is done through adding PSFs scaled by given fluxes into a map with the same noise level as our observed maps, and detecting these artificial objects with the same iteration procedure used for source extraction in our 850\,$\micron$ science maps. We construct the true noise map using the jackknife resampling technique. In each of our two overdensity fields, the stacked scan maps are randomly divided into two equal groups and co-added to create two half-made mosaic maps. The two half-made maps subtract each other to obtain a source-subtracted noise map. Then the noise map is multiplied by a factor of two to match the noise level of our 850\,$\micron$ science maps. We point out that another means to match the true noise level often used in literature is to scale the two half-made maps by $\sqrt{t_{1}t_{2}}/(t_{1}+t_{2})$, where $t_{1}$ and $t_{2}$ are the noise-weighted exposure times.  We verify that the two methods yield maps having almost identical noise level.  By doing so, all celestial sources are removed and pure noises are left in the jackknife maps. We compare S/N distributions within the effective area between our 850\,$\micron$ science maps and the jackknife noise maps in Figure~\ref{fig:hist}. One can see that the jackknife maps perfectly follow a normal distribution of random noises, while the science maps exhibit two tails from the normal distribution --- the positive tail is contributed by celestial sources, and the negative tail is due to the negative troughs around bright sources induced by the matched-filtering process.

We take our empirical PSF to create artificial sources of fixed flux density randomly distributed in the jackknife true noise maps. Following \cite{Geach2017}, the number of sources at a given flux density follows an initial distribution described well with a Schechter function:
\begin{equation}
    \frac{dN}{dS} = (\frac{N_0}{S_0})\  (\frac{S}{S_0})^{-\gamma} \exp(-\frac{S}{S_0}),
    \label{eq:Scfunction}
\end{equation}
where $dN$ is the source number per square degree at the given flux density $S$ within the flux bin $dS$.  These parameters are given by $N_0 = 7180$\,deg$^{-2}$, $S_0 = 2.5$\,mJy and $\gamma = 1.5$. We simulate sources in a flux range of $1-20$\,mJy. We do not consider the sources brighter than 20\,mJy as this is out of the flux range of sources in our catalogues. The spatial distribution of these artificial sources is randomly set, and clustering which could increase the number counts is not taken into account \citep{Cowley2015, Bethermin2017} 
We perform the same source extraction procedure as done to the science maps to recover the artificial sources. We regard a source as recovered  if it is extracted within the half beam ($\sim 8\arcsec$) of previously-detected artificial sources injected into the jackknife noise maps. And we pick up the brightest one if there are more than one source located in one beam.  We do not impose the minimum flux ratio on the recovered sources. Since the recovered sources with an extremely high flux ratio (thus a high boosting factor) only occupy a tiny fraction of the total source number and will not affect our results.
We generate ten thousands of simulated maps and matched the input catalogues  and recovered catalogues. Finally we obtain the matched catalogues with millions of recovered sources which can help us to investigate the observational bias. 

\begin{figure}
    \includegraphics[width=\columnwidth]{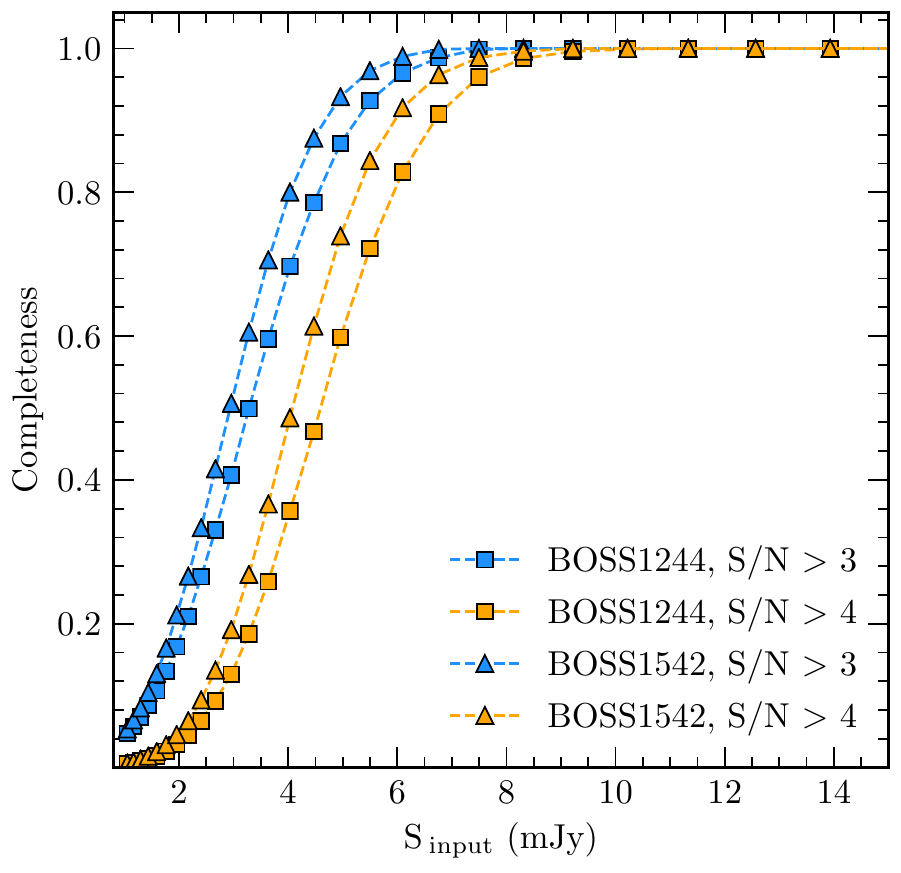}
    \caption{Completeness as a function of 850\,$\micron$ flux density within the area of S/N$>3$ (blue) and of S/N$>4$ (orange) in BOSS1244 (squares) and BOSS1542 (triangles). It is clear that the BOSS1244 map is slightly shallower than the BOSS1542 map and has a higher noise level. The completeness curve of BOSS1244 is shifted $\sim$ 0.5\,mJy toward the high end in comparison with that of BOSS1542.}
    \label{fig:comp}
\end{figure}

We define completeness as the ratio of the number of recovered sources to  the number of injected sources at fixed fluxes. Figure~\ref{fig:comp} shows the overall completeness as a function of intrinsic flux density in BOSS1244 and BOSS1542. Note that we only focus on the effective area where the noise level is less than 1.5 times the central noise. At $>$3$\sigma$, the completeness reaches 50\,per\,cent (80\,per\,cent) at the input flux density of 3.3\,mJy (4.5\,mJy) and 3.0\,mJy (4.0\,mJy) for BOSS1244 and BOSS1542, respectively. Furthermore, BOSS1244 (BOSS1542) achieves 50\,per\,cent completeness at 4.6 (4.1)\,mJy, and 80\,per\,cent completeness at 6.0 (5.3)\,mJy with a confidence of  $>$4$\sigma$. A relative lower completeness indicates a higher instrument noise in BOSS1244. 

Flux boosting factor refers to the ratio between the flux density evaluated from the recovered sources and their input flux density. One cause for this effect is the well-known Eddington bias \citep{Eddington1913}. Since the number of sources decreases exponentially from the faint end to the bright end, a source in a low flux bin  has a higher probability to be scattered up into higher flux bins due to the Gaussian noise fluctuation. On the other hand, the flux density of numerous sources below the detection limit can also contribute to the flux density of the detected sources. 
We utilize the recovered catalogues to estimate the deboosted (intrinsic) flux density for each source from the science maps.  Specifically, the observed flux density is taken from the distribution of the intrinsic flux density, $p(S_{\rm true})$, which can be simply estimated by the input flux density of injected sources within the observed flux density and local noise bins.  For each real source, we calculate the average input flux density of the injected sources within the bins of 0.5\,mJy centred on its observed flux and 0.2\,mJy centred on its local noise as the deboosted flux density of this source. Similarly, we obtain the uncertainty of the deboosted flux density by calculating the standard deviation of the input flux density within the specific bins above. The final science catalogues contain sources with the deboosted flux density  in the  range of $S_{850}=2.8-$16.6\,mJy. We use the deboosted flux density and the uncertainty of the detected sources to calculate the number counts in next section. 

\begin{figure}
    \includegraphics[width=\columnwidth]{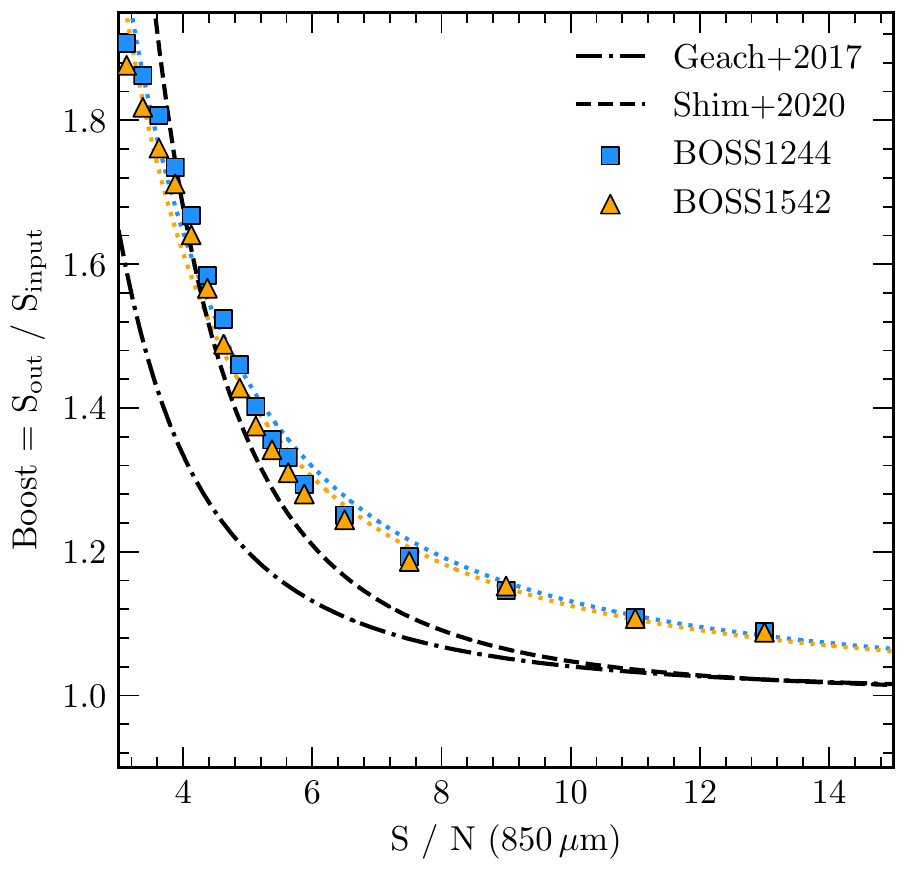}
    \caption{Boosting factor of  850\,$\micron$ source flux as a function of signal-to-noise ratio for BOSS1244 (squares) and BOSS1542 (triangles).  Here boosting factor measures the ratio of recovered flux to injected flux of a simulated 850\,$\micron$ source. The dotted lines represent the best-fitting power-law functions to the data points. For a comparison,  we plot the 850\,$\micron$ flux boosting curves from \protect \cite{Geach2017} with dot-dashed line and \protect \cite{Shim2020} with dashed line. Our 850\,$\micron$ flux boosting curve is apparently higher at the high-S/N end. We attribute this increase to the presence of an SMG overdensity in BOSS1244 and BOSS1542.}
    \label{fig:boost}
\end{figure}

\cite{Geach2017} found that the average boosting factor can be well described as a power-law function of S/N. We find a similar trend in our simulations, as shown in Figure~\ref{fig:boost}. The boosting factor in BOSS1244 is slightly higher than that in BOSS1542 in the low S/N regime, while the two are consistent to each other at high S/N. This small discrepancy in the low S/N regime is driven by the slightly shallower depth in BOSS1244. The flux density of our observations seem to be enhanced more when compared to the results of \cite{Geach2017} while less than that of \cite{Shim2020} in the low S/N regime. The boosting factor is higher than that given in literature at $>$6$\sigma$, which can also be  seen in the observations of GOODS-N (figure~11 in \citealt{Geach2017}) and MAMMOTH-1 (see the catalogue in \citealt{Arrigoni2018}). This is largely because bright sources are rare in the small coverage area, which are easily contaminated by the blended faint sources, leading to a higher boosting factor in the high S/N regime.

\subsection{Positional offset}

\begin{figure}
    \includegraphics[width=\columnwidth]{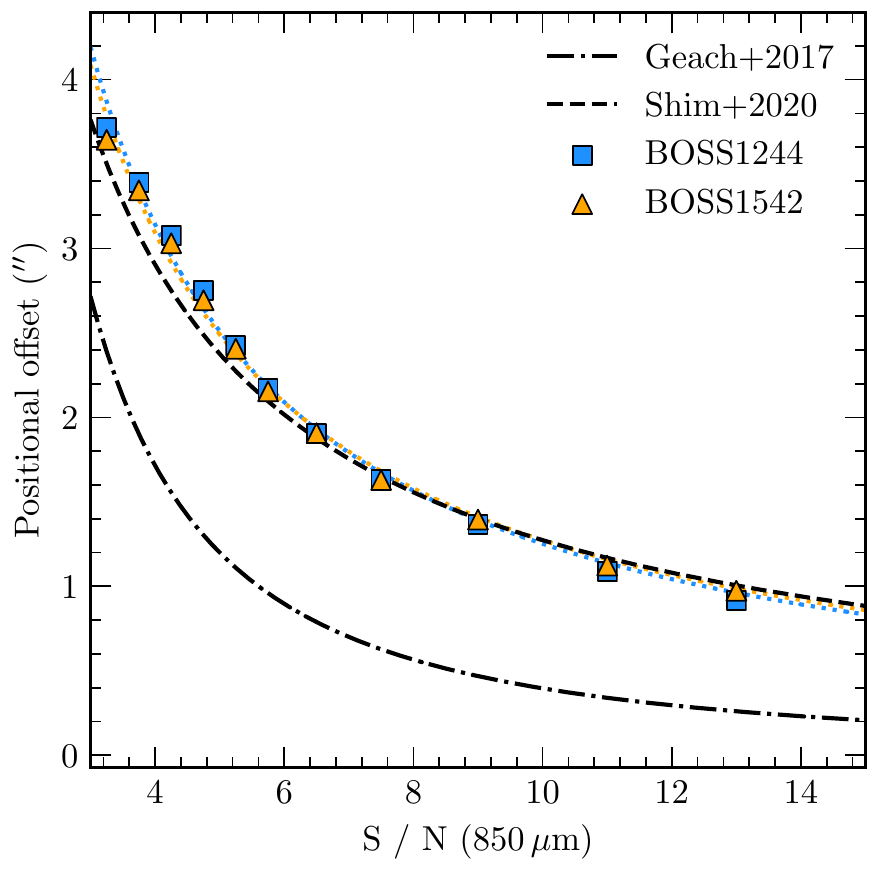}
    \caption{Positional offset of the injected and recovered 850\,$\micron$  sources in Monte Carlo simulations as a function of signal-to-noise ratio for BOSS1244 (blue) and BOSS1542 (orange). The dotted lines represent the power-law functions best-fitting the data points. The dashed line shows the result from \protect \cite{Shim2020}, in good agreement with ours. The dot-dashed line is the positional offset curve from the S2CLS survey \citep{Geach2017}, which has a higher accuracy in pinpointing the source locations.} 
    \label{fig:sep}
\end{figure}

In our Monte Carlo simulations, we match the input and output catalogues  and investigate the positional offset between the injected and recovered sources. The average offset as a function of S/N is shown in Figure~\ref{fig:sep}.  We estimate the positional offset at $4\sigma$ to be $3\farcs1$  and down to $\sim 2\arcsec$ at $6\sigma$ in both BOSS1244 and BOSS1542 with the pixel scale of $1\arcsec$ per\,pixel. The positional offset drops slowly to $1\arcsec$ at $12\sigma$. The positional offset as a functions of S/N was found to be well described by a single power-law function. We notice that our overall positional offset is slight higher than the result from the S2CLS survey \citep{Geach2017}, but almost identical to the power-law function given in \cite{Shim2020}. Our results of cross-matching for the optical and near-infrared counterparts will demonstrate that the position uncertainties are typically down to $3\arcsec$ at the $>$4$\sigma$ level. 

\subsection{False-detection rate}
\label{sec:fdr}

Although a jackknife noise map in principle gets rid of all signals from astronomical objects, spurious ``sources''  can be detected due to noise fluctuation.   For this reason we need to examine the false-detection rate (FDR) in our 850\,$\micron$ science maps.  FDR is defined as the ratio between the number of detections in the jackknife noise map and its science map \citep{Geach2017}. We use the jackknife method to generate 50 realizations in each field to eliminate bias caused by the random selection and applied source extraction to these maps as we did with the science maps.

\begin{figure}
    \includegraphics[width=\columnwidth]{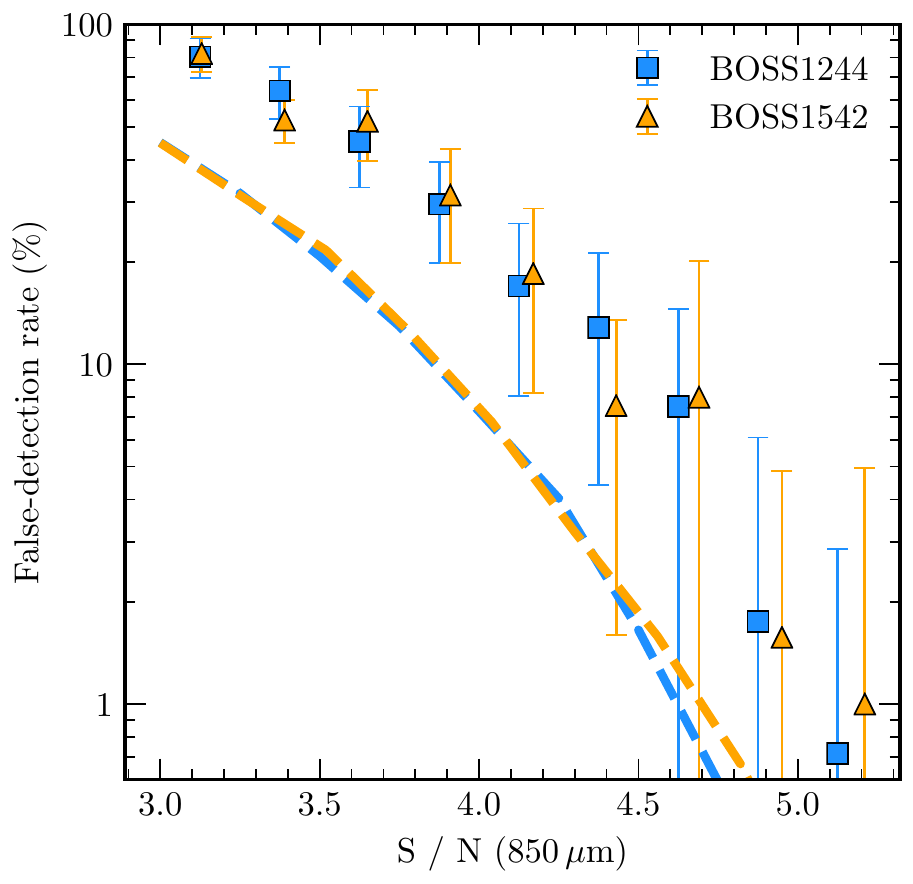}
    \caption{False-detection rate (FDR) as a function of signal-to-noise ratio of the 850\,$\micron$ science map for BOSS1244 (blue squares) and BOSS1542 (orange triangles).  The rate refers to the ratio of the average number of sources detected in 50 jackknife noise maps to the number of sources detected in the corresponding science map. Dashed lines demonstrate the cumulative FDR above a given S/N. The cumulative FDR is 7\,per\,cent at $>$4$\sigma$ and below 2\,per\,cent at $>$4.5$\sigma$.}    
    \label{fig:fdr}
\end{figure}

We estimate FDR as a function of S/N as well as the cumulative FDR above a given S/N. The results are shown in Figure~\ref{fig:fdr}. At $3\sigma$, we calculate the differential FDR to be 80\,per\,cent. The cumulative FDR is 45\,per\,cent at $>$3$\sigma$  for both of BOSS1244 and BOSS1542,  meaning that almost half objects in our preliminary catalogues  appear to be spurious.  When FDR decreases to 17 and 18\,per\,cent at $4\sigma$ in the two fields, the cumulative FDR drops to 7\,per\,cent at $>$4$\sigma$ and even down below 2\,per\,cent at $>$4.5$\sigma$ in both fields. We adopt the correction for FDR from  Figure~\ref{fig:fdr} in deriving number counts (next section). We refine the preliminary catalogues  with an S/N cut of $>$4$\sigma$  and obtain our final science catalogues. We estimate that there may be three and four spurious sources included in the science catalogues  of BOSS1244 and BOSS1542, respectively. Nevertheless, we keep this in mind when conducting studies of individual sources between $4\sigma$ and $5\sigma$. Observations with submillimetre interferometric telescopes (e.g., NOEMA and SMA) will help to verify the authenticity of these sources. We point out that the true point sources in deep 850\,$\micron$ observations of extragalactic fields are exclusively SMGs. We thus see the securely detected  850\,$\micron$ sources as SMGs in our two fields.

\section{RESULTS} \label{sec:results}

\subsection{Number counts} \label{sec:counts}


\begin{figure*}
     \includegraphics[width=\columnwidth]{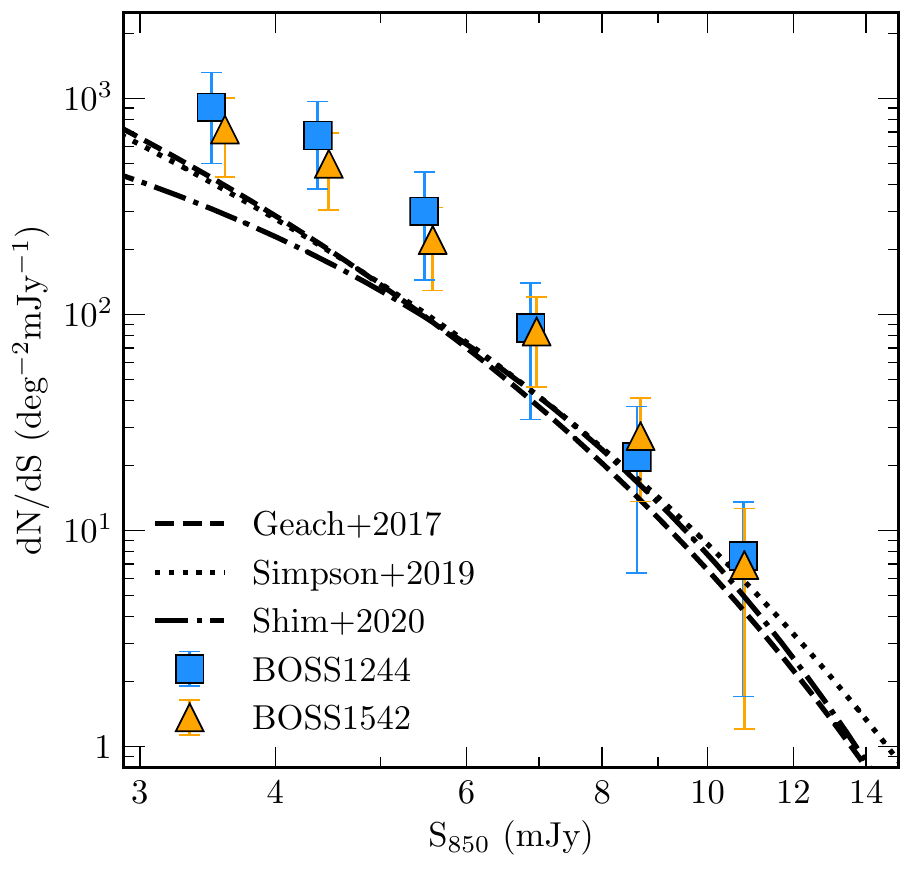}
     \includegraphics[width=\columnwidth]{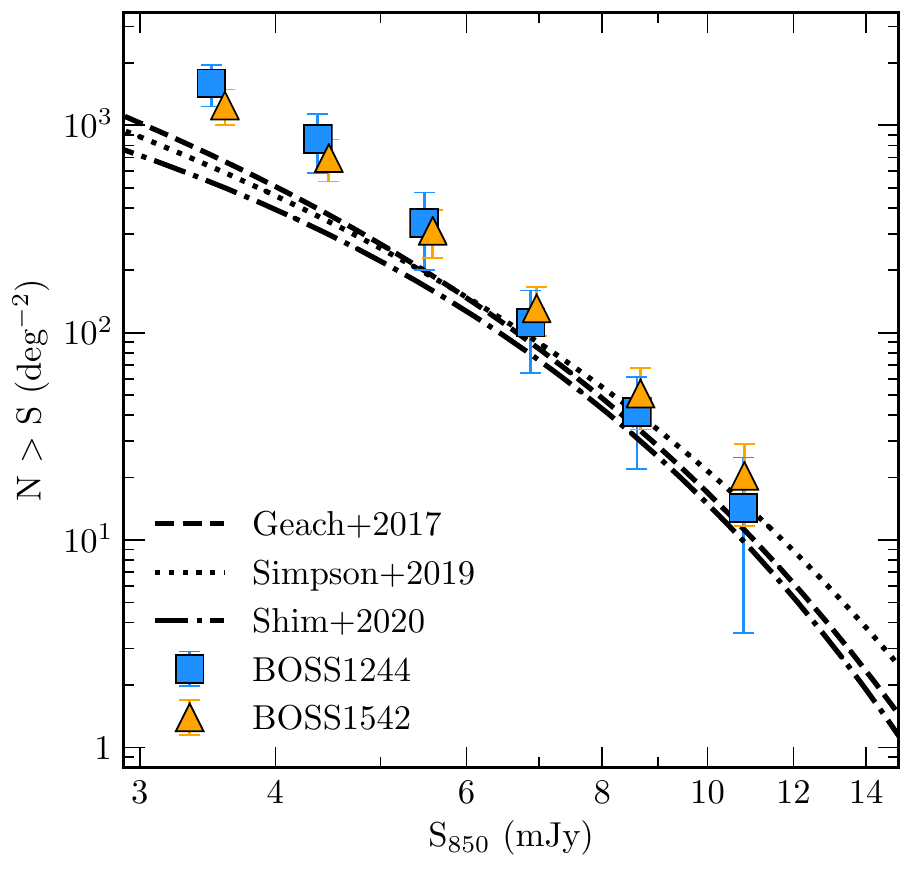}
     \caption{Differential number count (left) and cumulative number count (right) as a function of intrinsic 850\,$\micron$   flux density in  BOSS1244 and BOSS1542. The number counts are derived in the flux range of $S_{850}=4$--$12$\,mJy. We also plot the 850\,$\micron$ number counts of blank fields for comparison. Note that only \protect \cite{Geach2017} and \protect \cite{Simpson2019} gave the best-fitting parameters of the differential number counts. The other curves are fitted to the data points using a Schechter function. For BOSS1244 and BOSS1542, the differential number counts are higher than blank fields especially at fainter flux density ($S_{850}=4$--$8$\,mJy) and the cumulative number counts also larger than that of the general surveys at $S_{850}>4$\,mJy, indicating that BOSS1244 and BOSS1542 are both SMGs overdensities at $z \sim 2$.}
     \label{fig:counts}
\end{figure*}

The number density of SMGs provides important constraints on galaxy evolution models. We want to examine whether there is an excess of SMGs in our two MAMMOTH fields compared to blank fields, as these extreme overdensities may exhibit more galaxy interactions (e.g. merger) or gas reservoirs.  
To quantify the excess, we derive 850\,$\micron$ number counts using our samples of 43 (54) SMGs ($\geq$4$\sigma$) in BOSS1244 (BOSS1542). The 850\,$\micron$ fluxes in our catalogues  have been corrected for the boosting effect given in Figure~\ref{fig:boost}. Still, corrections for FDR given in Figure~\ref{fig:fdr} and completeness given in Figure~\ref{fig:comp} are required to obtain the true number counts. For each source, a conversion number of $ (1-\mathcal{F}_{\rm i})\ /\ \mathcal{C}_{\rm i}$ is applied, where $\mathcal{F}$ and $\mathcal{C}$ represent FDR and completeness at the deboosted flux and S/N of the source, respectively. The number density is calculated from the sum of the conversion numbers of all sources within a give flux bin averaged over the effective area and the bin width of each source.  
The fluxes of our detected sources range from  $S_{850}=4$\,mJy to $S_{850}=12$\,mJy. Here we ignore one exceptionally-bright source ($S_{850}=18$\,mJy) in BOSS1542. We split the flux range into six even bins in the logarithm space to ensure that each bin contains enough sources to reduce the shot noise. The flux ranges are slightly different between the two fields. Doing so we obtain the differential and cumulative number counts of  850\,$\micron$-detected SMGs in our two MAMMOTH fields.   We construct the number counts to the faint end at about 3\,mJy, reaching the lower limit of the deboosted fluxes of our source catalogs.  

The errors of number counts are estimated from rebuilding the number counts with the mock fluxes fluctuating within uncertainties. The error of each source's deboosted flux  obeys a normal distribution (Figure~\ref{fig:boost}).  A set of `new'  intrinsic fluxes can be obtained by adding randomly-generated errors to the intrinsic fluxes and re-calculate the differential number counts. This procedure is repeated 1000 times for a statistics. The mean values are taken as the final number counts, and the dispersion are taken as their errors. We also construct the cumulative number counts following the same procedure described above. We list the number counts for BOSS1244 and BOSS1542 in Table~\ref{tab:tab2}, and show them in Figure~\ref{fig:counts}. We take the widest 850\,$\micron$ survey of S2CLS over $\sim 5$\,deg$^2$ \citep{Geach2017},  the SCUBA-2 COSMOS survey \citep[S2COSMOS;][]{Simpson2019},  
and the survey of the extended North Ecliptic Pole SCUBA-2 (NEPSC2) region over 2\,deg$^2$ \citep{Shim2020} as for the blank fields and plot the number counts from these surveys for comparison. The best-fitting functions to the differential number counts in S2CLS and S2COSMOS are available in literature.  We best fit the differential number counts in NEPSC2 by using the published data in \citet{Shim2020}.  The Schechter function is used to best fit the cumulative number counts in all three blank fields. 

\begin{table}
\centering
\caption{The SCUBA-2 850\,$\micron$ differential and cumulative number counts in BOSS1244 and BOSS1542.} \label{tab:tab2}
\renewcommand{\arraystretch}{1.3}
\begin{tabular}{cccc}
    \hline    \hline
    Field &    $S_{850}$    &    $dN/dS$    &    $N$ (>$S$) \\ [3pt]
    \hline
                 	&3.5&	909.6 $\pm$ 410.1&		1593.1 $\pm$ 359.4 \\ [2pt]
                		&4.4&	673.9 $\pm$ 293.7&		859.2 $\ \pm$ 271.0 \\ [2pt]
                		&5.5&	300.3 $\pm$ 156.1&		337.4 $\ \pm$ 136.8 \\ [2pt]
BOSS1244    	&6.9&	86.4 $\ \pm\ $ 53.7&		111.9 $\ \pm\ $  47.8 \\ [2pt]
                		&8.6&	21.9 $\ \pm\ $ 15.5&		  41.5 $\ \ \pm\ $ 19.5 \\ [2pt]
                		&10.8&	7.6 $\ \ \pm\ \ \ $ 5.9&	  14.3 $\ \ \pm\ $ 10.7 \\ [2pt]
    \hline
                 	&3.6&	718.4 $\pm$ 286.4&	1246.0 $\pm$ 242.5 \\ [2pt]
                		&4.5&	498.4 $\pm$ 193.7& 695.1 $\pm$ 158.8 \\ [2pt]
                		&5.6&	221.2 $\pm\ $ 91.9&	310.1 $\pm\ $ 80.9 \\ [2pt]
BOSS1542    	&7.0&	83.3 $\ \pm\ $ 37.2&	131.0 $\pm\ $ 35.0 \\ [2pt]
                		&8.7&	27.4 $\ \pm\ $ 13.7&	50.9 $\ \pm\ $ 16.8 \\ [2pt]
                		&10.8&	6.9 $\ \ \pm\ \ $ 5.7&	20.4 $\ \pm\ \ $ 8.7 \\ [2pt]
    \hline
\end{tabular}
\end{table}

\begin{table}
\centering
\caption{ Comparison of our 850\,$\micron$ number counts in BOSS1244 and BOSS1542 with those of three surveys of blank fields. }
\label{tab:tab3}
\renewcommand{\arraystretch}{1.3}
\begin{tabular}{llrr}
    \hline    \hline
  &  Number Count  &  $N_0^{\rm fit}$ &   Ratio \\ [3pt]
    \hline
    			 &  \citet{Geach2017} &     15672 $\pm$ 2330&       2.2 $\pm$ 0.5  \\ [2pt]
 BOSS1244       &  \citet{Simpson2019}&   9030 $\pm$ 1830&	  1.8 $\pm$ 0.6  \\ [2pt]
			 &   \citet{Shim2020} &      8119 $\pm$ 1660&	  1.9 $\pm$ 0.4 \\ [2pt]
			 &  Mean  & \multicolumn{2}{c}{$2.0 \pm 0.3$} \\ [2pt]
			 &  Density peak (r = 1\,Mpc) & \multicolumn{2}{c}{$5.1 \pm 0.9$} \\ [2pt]
    \hline
    			 & \citet{Geach2017}&     15784 $\pm$ 958\,\,\,&       2.2 $\pm$ 0.4  \\ [2pt]
  BOSS1542  	 & \citet{Simpson2019}&   9555 $\pm$ 1120&	  1.9 $\pm$ 0.2  \\ [2pt]
   			 &  \citet{Shim2020}&      8577 $\pm$ 989\,\,\,&	  2.0 $\pm$ 0.2 \\ [2pt]
  			 &  Mean  & \multicolumn{2}{c}{$2.1 \pm 0.2$} \\ [2pt]
			 &  Density peak (r = 1\,Mpc)  & \multicolumn{2}{c}{$4.3 \pm 0.5$} \\ [2pt]
    \hline
\end{tabular}
\end{table}

We note that both of BOSS1244 and BOSS1542 are  SMG overdensities compared to general fields. The conversion number described above could be biased by the fiducial model we adopted and thus affect the derived number counts. Following \cite{Nowotka2022}, we also construct the intrinsic number counts using the self-consistent method  in BOSS1244 and BOSS1542 to test the results above. We firstly obtain the raw number counts by calculating the difference between the source counts constructed from science maps and jackknife maps. The raw number counts are best fitted with the Schechter function using the least-squares method. Similar to the simulations above, we inject artificial sources into the jackknife maps with the best-fitting model and extract sources as before. The corrected counts are obtained by scaling the raw counts with the ratios between the input model and recovered counts. Then the best fit of the recovered counts are used as the input model in the next iteration. The parameter $S_0$ is fixed as 2.5 during the iterations and will not affect the results. The iterations are stopped when the recovered counts are consistent with the raw counts within the uncertainties, and the final input model is regarded as the intrinsic number counts of the two protoclusters. We find that the intrinsic models successfully reproduce the raw number counts and agree well with the number counts constructed from conversion number of the sources within the uncertainties.  

The beam size of SCUBA-2 at 850\,$\micron$ is relatively large ($\sim15\arcsec$) and  leads  some  sumillimetre sources to be composed of multiple components blended into a single beam \citep{Wang2011, Karim2013, Hodge2013}.  This issue could affect the determination of number counts. Studies based on submillimetre interferometric observations revealed the fraction of single-dish detections to include multiple components  increases with source flux \citep{Stach2018}. The number counts derived from ALMA observations were found to be slightly lower than those of SCUBA-2 surveys  \citep{Simpson2015, Stach2018, Hill2018}. The majority of SMGs of multiplicity are found to be projections of sources at different redshifts \citep{Hayward2018}, while the the physically-associated multiple components are also identified for some of brightest SMGs \citep[e.g.][]{Fu2013}. Since the number counts in BOSS1244 and BOSS1542 could be affected by multiplicity in a similar way, the changes in our estimates of the overdensity factor are expected to be negligible. 

\subsection{SMG overdensities}
\label{sec:overdensity}

BOSS1244 and BOSS1542 are  overdensities traced by H$\alpha$ emitter with an overdensity factor ($ \delta_{\rm gal} = (N_{\rm gal} - N_{\rm field})/N_{\rm field}$) of $5.6\pm0.3$ and $4.9\pm0.3$ in a volume of $ 54 \times 32 \times 32h^{-1}\,{\rm cMpc}^3$, respectively \citep{Zheng2021}.  Their number counts at 850\,$\micron$  can be compared with those of blank fields to see whether these two protoclusters are also SMG overdensities. As shown in Figure~\ref{fig:counts}, the differential number counts in BOSS1244 and BOSS1542 are both higher than those of the blank fields (S2CLS, S2COSMOS and NEPSC2) over the flux  range of $S_{850}=4-$12\,mJy.  We notice that the excess of our number counts at $S_{850}>8$\,mJy appear  smaller than that at the faint end, indicating that the excess of SMGs in our two overdensity fields is mostly contributed by the faint SMGs of $S_{850}<8$\,mJy.  
Our analyses do not involve the bright end ($S_{850}>12$\,mJy), where the number counts still have large uncertainties limited by the sample sizes \citep{Shim2020}. Another uncertainty comes from the faint-end completeness correction. We notice that the number counts in the lowest bins appear to be lower than the extrapolation from the best-fitting of other data points. This deficiency is likely due to the insufficient correction for the incompleteness at the faintest end,  and induced by the combination of the S/N limitation and the uncertainty in deboosting fluxes.  From the cumulative number counts, we see that BOSS1244 contains clearly more SMGs than that of blank fields particularly at the low-flux end ($S_{850}=4-$6\,mJy).  We conclude that  both of our two MAMMOTH fields host an excess of SMGs  in compared with the blank fields.  Considering the insufficient completeness correction at the faintest end, the excess in the number counts for our sample might be even larger.

\begin{figure*}
     \includegraphics[width=\columnwidth]{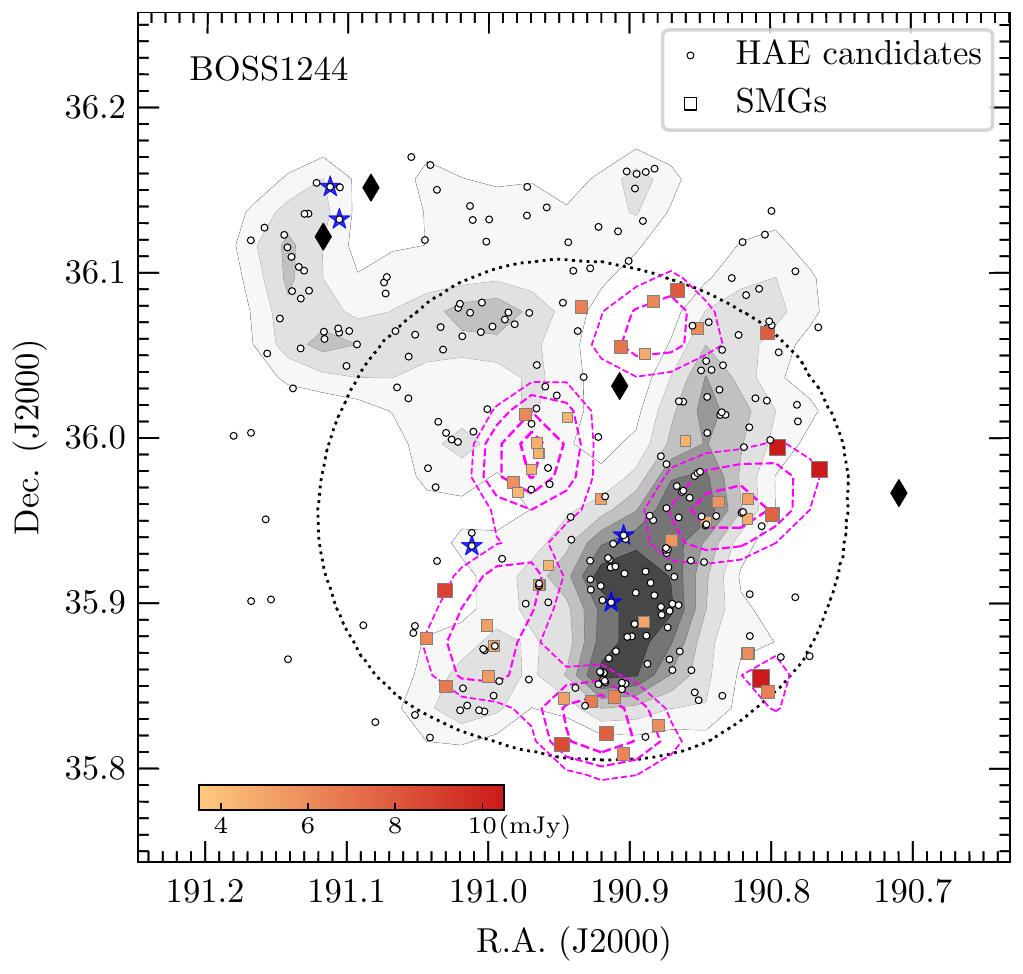}
     \includegraphics[width=\columnwidth]{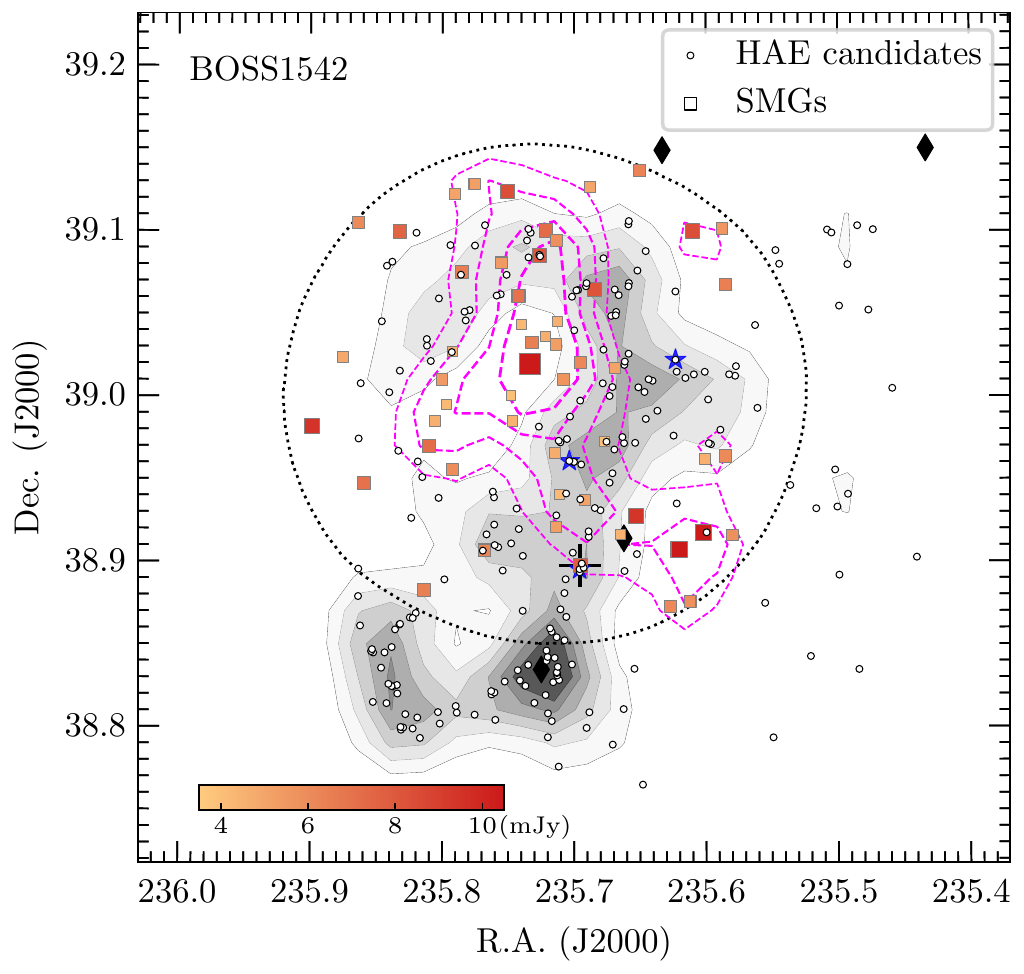}
     \caption{Comparing spatial distributions of 850\,$\micron$ sources with HAE candidates  in BOSS1244 (left) and BOSS1542 (right). The HAE candidates are shown with open circles against their density maps and the filled grey contours ([4, 8, 12, 16, 20, 24] $\times$ 0.071\,arcmin$^{-2}$) adopted from \protect \cite{Zheng2021}.  
     The magenta-dashed lines show the contours of the surface number density of SMGs smoothed by a Gaussian kernel of $\sigma=1.2$\,arcmin ($1.9$\,cMpc at $z = 2.24$). The contour levels refer to 1.6, 2.4, 3.2 and 4 times the SMG number density in the general fields ($\sim400$\,deg$^{-2}$). Black diamonds mark the coherently groups of strong Ly$\alpha$ absorption systems and blue stars represent the quasars at $z\sim2.24$. The 850\,$\micron$ sources are shown with squares colour-coded and size-scaled by their observed fluxes.  Black dotted curves (nearly a circle with a diameter of $9.2$\,arcmin) enclose the effective area of our deep SCUBA-2 observations. The black cross indicates the spectroscopically confirmed quasar which might be counterpart for SMG 1542.07 (see Section~\ref{sec:counterpart}). } 
     \label{fig:density}
\end{figure*}
We further quantify the  excess of SMGs  in BOSS1244 and BOSS1542 using the method made use of the differential number counts given in \cite{Arrigoni2018}. We take S2CLS, S2COSMOS and NEPSC2 as representative of blank fields. The differential number counts of the three surveys are well constrained and best fitted with the Schechter function given in Equation~\ref{eq:Scfunction}.   The best-fitting parameters  from S2CLS \citep{Geach2017}  give $N_0 = 7180\pm 1220$\,deg$^{-2}$, $S_0 = 2.5\pm 0.4$\,mJy and $\gamma = 1.5\pm0.4$. \cite{Simpson2019} reported the best-fitting parameters  from S2COSMOS as $N_0 = 5.0^{+1.3}_{-1.4}\times 10^3$\,deg$^{-2}$, $S_0 = 3.0^{+0.6}_{-0.5}$\,mJy and $\gamma = 1.6^{+0.3}_{-0.4}$.  The data points of the number counts from the NEPSC2 survey are available in \cite{Shim2020}. We fit a Schechter function to these data points and obtain the best-fitting parameters as $N_0 = 4191\pm 144$\,deg$^{-2}$, $S_0 = 1.9\pm 0.1$\,mJy and $\gamma=0.2\pm 0.1$. For the differential number counts in BOSS1244 and BOSS1542, we use the Schechter function of fixed $S_0$ and $\gamma$ from one of the above three best-fitting functions  to fit the data points and obtain the best-fitting $N_0$.  In this way, we can fairly compare our best-fitting $N_0$ with that from given surveys and ascertain the the excess of SMGs in for the two MAMMOTH fields. Table~\ref{tab:tab3} presents our best-fitting $N_0$ to the number counts of BOSS1244 and BOSS1542  and the ratios of our $N_0$ to those from the three surveys. The mean of the three ratios is adopted to measure the overdensity of SMGs in BOSS1244 and BOSS1542, giving $2.0\pm0.3$ and $2.1\pm0.2$ times the number density of SMGs in blank fields.  We note that the $N_0$ ratio represents the overall scaling factor between these number counts, accounting for the discrepancies over the entire flux range examined.

The cumulative number counts are dominated by the number density of  submillimetre sources at the faint end. We compare the cumulative number counts between our surveys and the three reference surveys.  From \cite{Geach2017} and \cite{Shim2020},  one can obtain the cumulative number counts in the S2CLS and NEPSC2 surveys as 508.0\,deg$^{-2}$ and 391.5\,deg$^{-2}$ at $S_{850}\geq 4.0$\,mJy with a similar survey depth.  At the same flux cut the number density of SMGs is 1138.8\,deg$^{-2}$ and 965.6\,deg$^{-2}$ for BOSS1244 and BOSS1542, respectively.   At $S_{850}\geq 4.2$\,mJy, we obtain 988.0\,deg$^{-2}$ and 838.2 deg$^{-2}$ for BOSS1244 and BOSS1542 in comparison with 398\,deg$^{-2}$ in S2COSMOS \citep{Simpson2019}, respectively.  We estimate the mean number density of SMGs to be $2.5\pm0.3$ ($2.1\pm0.2$) for BOSS1244 (BOSS1542)  relative to that of the blank fields, in good agreement with our results based on the fitting of the differential number counts. We remind that our number counts might be underestimated at the faintest bin (mentioned in the beginning of this Section) and thus slightly bias the cumulative number counts. We thus adopt $2.0\pm0.3$ ($2.1\pm0.2$) as the number density of SMGs in BOSS1244 (BOSS1542)  comparing to the blank fields.

The number density of SMGs in either BOSS1244 or BOSS1542 is significantly higher than that of the blank fields. It is reasonable to attribute the excess of SMGs to the overdensity at $z=2.24$.  This has been successfully verified with H$\alpha$ emitters  \citep{Zheng2021,Shi2021}.  We link the excess of SMGs in our two MAMMOTH fields to the presence of 850\,$\micron$-detected SMG overdensity at $z=2.24$. There are 43 (54) SMGs detected over an effective area of 264\,arcmin$^2$  in the deep 850\,$\micron$ map of BOSS1244 (BOSS1542).  Considering the redshift distribution, we calculate the fraction of SMGs  located in $2.2<z<2.3$ to be 5.7\,per\,cent comparing to the whole redshift range in the AS2UDS survey \citep{Dudzeviciute2020}, which has a survey area of $\sim1$\,deg$^{2}$ and make the cosmic variance negligible \citep{Simpson2019}. Then we obtain that the volume density of SMGs in this redshift slice is $2.6 \times 10^{-5}$\,cMpc$^{-3}$ at $S_{850}\geq 4.0$\,mJy in the blank fields. If we simply assume the two overdensities span a narrow redshift range ($\delta z<0.042$ given by the band width of H$_2$S(1) filter), the volume of the redshift slice of $z = 2.246\pm0.021$ over 264\,arcmin$^2$ corresponds to a co-moving box of $54.3 \times 665.6$ (=36142)\,h$^{-3}$\,cMpc$^3$. If the excess of SMGs (i.e., $(1.8 - 1)/1.8 \times 54$ for BOSS1542) are associated with the $z=2.24$ overdensities, we obtain that SMGs in the two overdensities may reach a volume density of $\sim(40$--$70)\times 10^{-5}$\,cMpc$^{-3}$, 15$-$30 times that of the blank fields. After enlarging the redshift slice of the two protoclusters to that of blank field ($\delta z=0.1$), we still obtain the volume density of SMG to be 6--11 times that of the blank fields.
The overdensity factor of SMGs, although with large uncertainties,  is even higher than that of HAEs  in these two protoclusters \citep{Shi2021}, indicating that there are more extreme starbursts with enhanced star formation in respect to normal star-forming galaxies (i.e., HAEs) in the two $z=2.24$ protoclusters.

Additionally, we estimate the overdensities and SFR densities in the Mpc-scale regions at the density peaks of SMGs in the two protoclusters. We focus on the most dense regions located in the valley between the NE and SW components in BOSS1244 and the void of the filament in BOSS1542. There are seven and nine submillimetre sources within a radius of 1\,Mpc in these two density peaks. Using the method above we obtain the overdensity is $5.1\pm0.9$ and $4.3\pm0.5$ that of general fields  in the density peaks of BOSS1244 and BOSS1542 respectively, which is also listed in Table~\ref{tab:tab3}. We estimate SFRs from 850\,$\micron$ fluxes using the formula from \cite{Cowie2017} as 
\begin{equation}
    {\rm SFR}\ ({\rm M}_\odot\,\rm{yr}^{-1}) = (143\pm20) \times S_{850\,\micron}\,(\rm{mJy}), 
    \label{eq:SFR}
\end{equation}
assuming a \citet[]{Kroupa2001} initial mass function. We then calculate the SFR densities within a sphere of 1\,Mpc radius to be $1600\pm630$ and $1500\pm550$\,M$_{\odot}$\,yr$^{-1}$\,Mpc$^{-3}$ in the SMG density peaks in BOSS1244 and BOSS1542.  Such high SFR densities are consistent with those in literature \citep{Clements2014, Dannerbauer2014, Kato2016, Arrigoni2018, Nowotka2022}, almost four magnitudes greater than the cosmic mean density \citep{Madau2014}.  This result indicates that there exist similar star formation activities in the off-peak regions of the two overdensities comparing to other protoclusters and ELANs. 

\begin{figure}
    \includegraphics[width=\columnwidth]{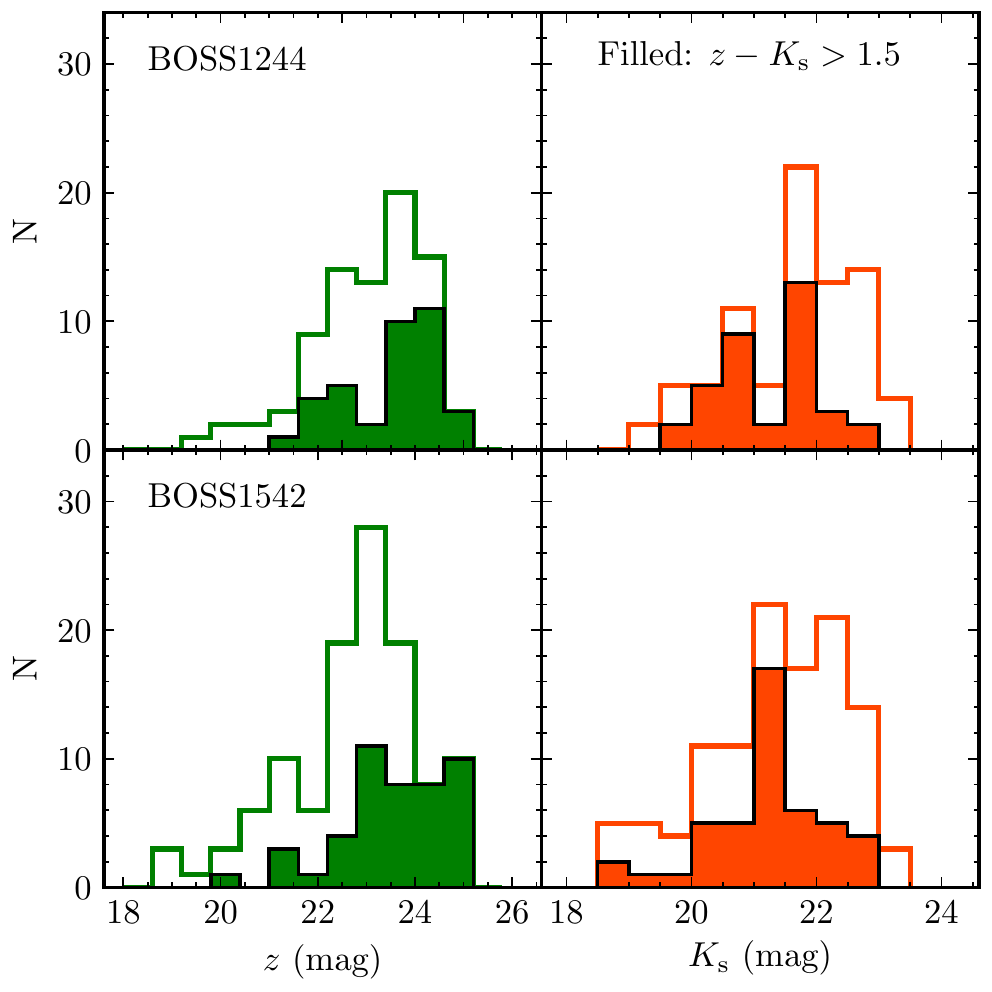}
    \caption{The $z$ (green) and $K_{\rm s}$ (red) magnitude distributions of 82 and 113 objects detected in both $z$ and $K_{\rm s}$ bands within the SCUBA-2 850\,$\micron$ beam size for 43 and 54 SMGs in BOSS1244 (top) and BOSS1542 (bottom), respectively.  The filled histograms represent the objects with $z-K_{\rm s}>1.5$, which are more likely to be the counterparts of SMGs.}
    \label{fig:color}
\end{figure}

\begin{figure*}
    \includegraphics[width=\columnwidth]{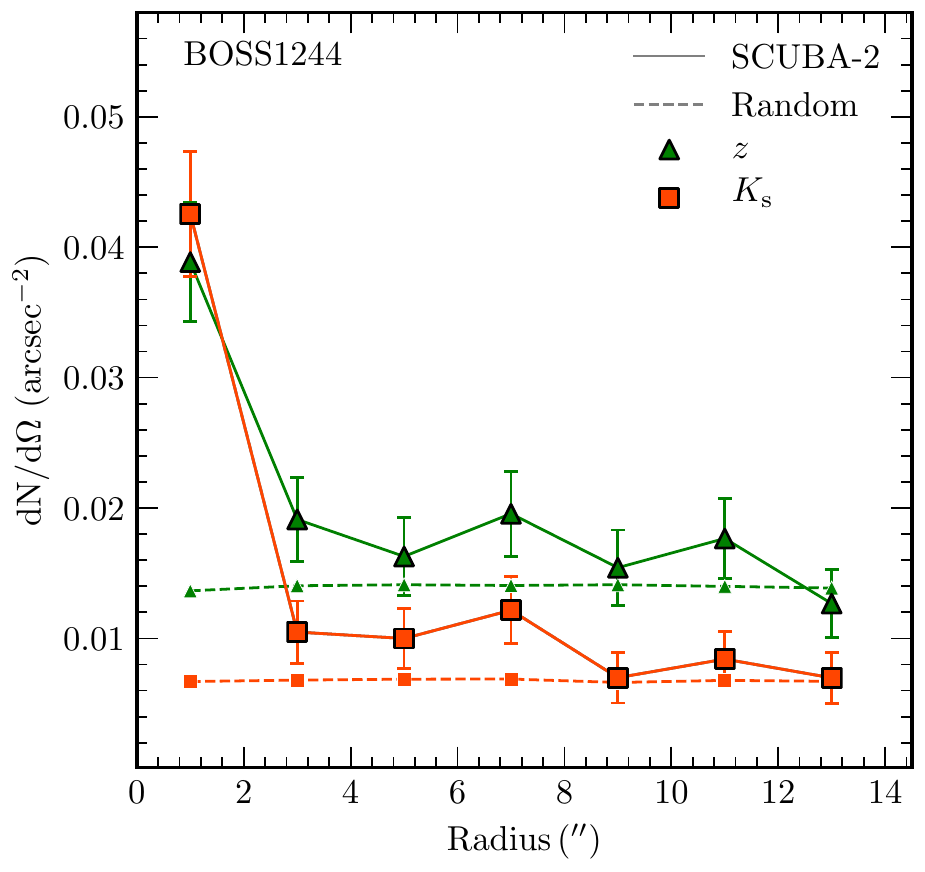}
    \includegraphics[width=\columnwidth]{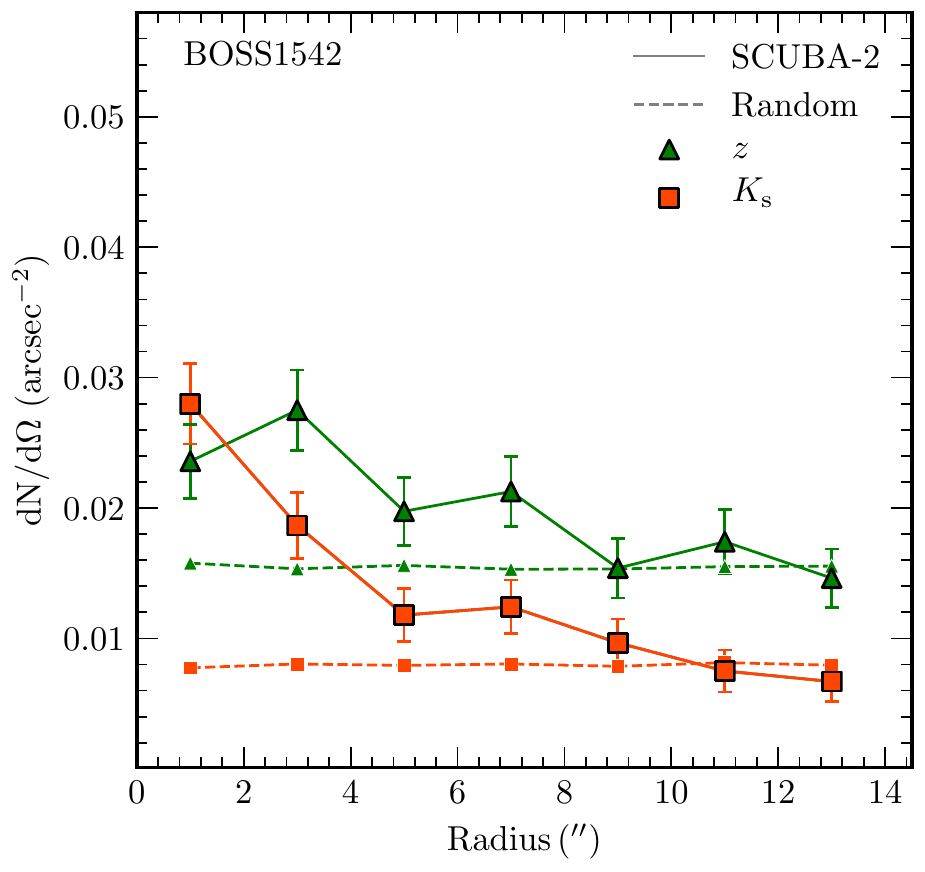}
    \caption{The surface density of the $z$ and $K_{\rm s}$ selected galaxies around the SCUBA-2 detections and the random positions as a function of radius in BOSS1244 (left) and BOSS1542 (right). The $z$ and $K_{\rm s}$ a distinguished by magenta and red symbols. The solid lines show the positions centred on the SCUBA-2 detections and the dashed liens are calculate from the average number density of 10\,000 random positions in two fields.}
    \label{fig:radius}
\end{figure*}

\subsection{Spatial distributions of SMGs}
\label{sec:densitymaps}

It has been shown in \cite{Zheng2021} that BOSS1244 and BOSS1542 are two massive structures of HAEs over scales of 32\,$h^{-1}$\,comoving\,Mpc (cMpc), having an overdensity factor of $5.6\pm0.3$ and $4.9\pm0.3$, respectively.  When focusing on the central high-density regions, the overdensity factor may increase by a factor of $\sim 2-4$, and thus to be the most extremely galaxy overdensities at $z>2$ discovered to date. The followup near-infrared (NIR) spectroscopic study of their HAE candidates confirmed them to be extreme protoclusters of galaxies  at $z=2.24$ \citep{Shi2021}. In BOSS1244, the HAE density map consists of two components: one extended low-density North-East (NE) and the other elongated high-density South-West (SW) component. The latter component contains two distinct components separated from each other along the line of sight.  BOSS1542 is spectroscopically confirmed to be a giant filamentary structure at $z = 2.241$. 

We examine the spatial distribution of SMGs in BOSS1244 and BOSS1542, in comparison with the spatial distribution of HAEs. We adopt the density maps of HAEs and the same contours from \cite{Zheng2021}, as shown in Figure~\ref{fig:density}.  The coherently strong Ly$\alpha$ absorption systems (CoSLAs), which footprint on the spectra of background quasars and are used to probe these overdensities,  are marked as black diamonds. Blue stars represent the quasars  belonging to the two overdensities and  included in the HAE sample. The dotted curves enclose the effective area of our SCUBA-2 850\,$\micron$ science maps, being smaller than the $20\arcmin \times 20\arcmin$ coverage of deep narrow and broad $K_{\rm s}$-band observations for HAEs.  The HAE candidates are shown with circles and SMGs ($\geq$4$\sigma$) are shown with squares sized and colour coded by their 850\,$\micron$ fluxes. Note that these HAE candidates contain $\sim$80\,per\,cent  true HAEs and the density maps are not biased by non-HAEs which are believed to be randomly located as fore- or background sources \citep[see][for more details]{Zheng2021,Shi2021}. Similarly, we build density maps and contours for SMGs. We divide the effective area of our 850\,$\micron$ maps into a grid of $1\arcmin\times1\arcmin$ cells,  calculate the number density of SMGs within each cell, and smooth the density map with a Gaussian kernel of $\sigma=1.5$\,arcmin.  Our samples of 850\,$\micron$ sources consist of SMGs belonged to the $z=2.24$ overdensities as well as fore- or background SMGs.  We take 400\,deg$^{-2}$ as the number density of SMGs down to $S_{850}=4$\,mJy  in the general fields and set the contour levels as 2, 3, 4 and 5 times the general-field density.  The magenta long-dashed curves  Figure~\ref{fig:density} show the density contours of SMGs in BOSS1244 and BOSS1542. 

From Figure~\ref{fig:density} one can clearly see that SMGs do not spatially concentrate on the densest regions traced by HAEs in both BOSS1244 and BOSS1542. Instead, they locate mostly around the HAE density peak regions.  Few of our sample SMGs are seen in these density peak regions. Most of the SMGs in BOSS1244 are detected in the valley between the  NE and SW components, as well as the low-density regions surrounding the SW component.  No SMG is detected inside of the NE component although only half of the NE area is effectively covered by our deep SCUBA-2 observations.   In BOSS1542, SMGs are located preferentially off the ridge regions of the filamentary structure.  Several SMGs seen in the density peak area might be projected from the outskirts of the dense structure along the line of sight. 
We use the two-dimensional Kolmogorov-Smirnov test to quantify the difference between the distributions of SMGs and HAEs in our two protoclusters \citep{Peacock1983, Fasano1987}. We calculate the observed D-statistic between the coordinates of SMGs and HAEs is 0.285 (0.398) in BOSS1244 (BOSS1542). The significance level is 0.0177 and 0.0004 in BOSS1244 and BOSS1542, meaning that the SMGs and HAEs are drawn from different spatial distributions at a confidence of 2.4$\sigma$ and 4$\sigma$, respectively. 

We notice that the brightest SMG with a flux density of $S_{850}=18.1$\,mJy (deboosted to $S_{850}=16.6$\,mJy) in our samples resides in a blank (HAE-free) region half encircled by the filamentary structure.  Interestingly, about ten SMGs are clustered around this source within 5\,arcmin$^2$ and form an extreme SMG overdensity.  The nature of this SMG overdensity and how it formed remain to be explored.   Followup spectroscopic observations are strongly demanded to study the member galaxies and connections with the surrounding filamentary structure in BOSS1542. In brief, \textit{a prominent discovery from the spatial distributions of our sample SMGs  is that they are mostly located around the density peak regions of HAEs in both BOSS1244 and BOSS1542, and barely occur in these peak regions.}  

\subsection{Optical/NIR counterparts}
\label{sec:counterpart}

 Due to the coarse resolution of  SCUBA-2 and the lack of the high-resolution interferometric data, it is  impossible to precisely identify the optical/NIR counterparts for SCUBA-2 850\,$\micron$ sources  in the two fields. On the other hand, we can examine the optical/NIR counterparts of the SMGs in a statistic manner.  The optical-to-NIR colour helps to pick up the SMG counterparts because SMGs are heavily  obscured by dust and usually red in the rest-frame UV-optical colour \citep{Chen2016}.  It is worth noting that the optical/NIR counterparts could lie $1-8\arcsec$ off from the nominal position of SCUBA-2 detections, while the majority of the counterparts have $z-K_{\rm s}$ colour larger than 1.5 \citep[see figure~4 of][]{An2018}. So we choose $z-K_{\rm s}>1.5$ as the criterion for counterparts to examine the likelihood of the optical/NIR objects within the SCUBA-2 beam. We obtained the $U$ and $z$-band  imaging data in the BOSS1244 field, and $U$, $B$, $V$ and $z$-band imaging data in the BOSS1542 field, using the Blue Camera on board the Large Binocular Telescope (LBT) in 2018 (Shi~D.~D. et al., in preparation).  The $z$ -band images of BOSS1244 (BOSS1542) reach a depth of 25.12 (25.39)\,mag above the  5$\sigma$ significance. Our $K_{\rm s}$ images  reach a 5$\sigma$ depth of 23.29 and 23.23\,mag for the two fields \citep[][]{Zheng2021}. We use SExtractor to construct source catalogues from the $z$ and $K_{\rm s}$-band images of BOSS1244 and BOSS1542.  We remove point sources and those brighter than 18\,mag that are usually low-$z$ galaxies. Then we select the $z$ and $K_{\rm s}$ sources around the positions of SMGs with a tolerance of $7\farcs5$  (i.e.,  the half of SCUBA-2 beam size at 850\,$\micron$). Doing so we obtain 82 and 113 of optical/NIR objects as potential counterparts for SMGs in BOSS1244 and BOSS1542. Of them,  36 and 46 satisfy $z-K_{\rm s}>1.5$, as shown in Figure~\ref{fig:color}.  We estimate about 40\,per\,cent of the optical/NIR objects within the beam size of SMGs are likely to be their counterparts.

\begin{figure*}
    \includegraphics[width=0.95\textwidth]{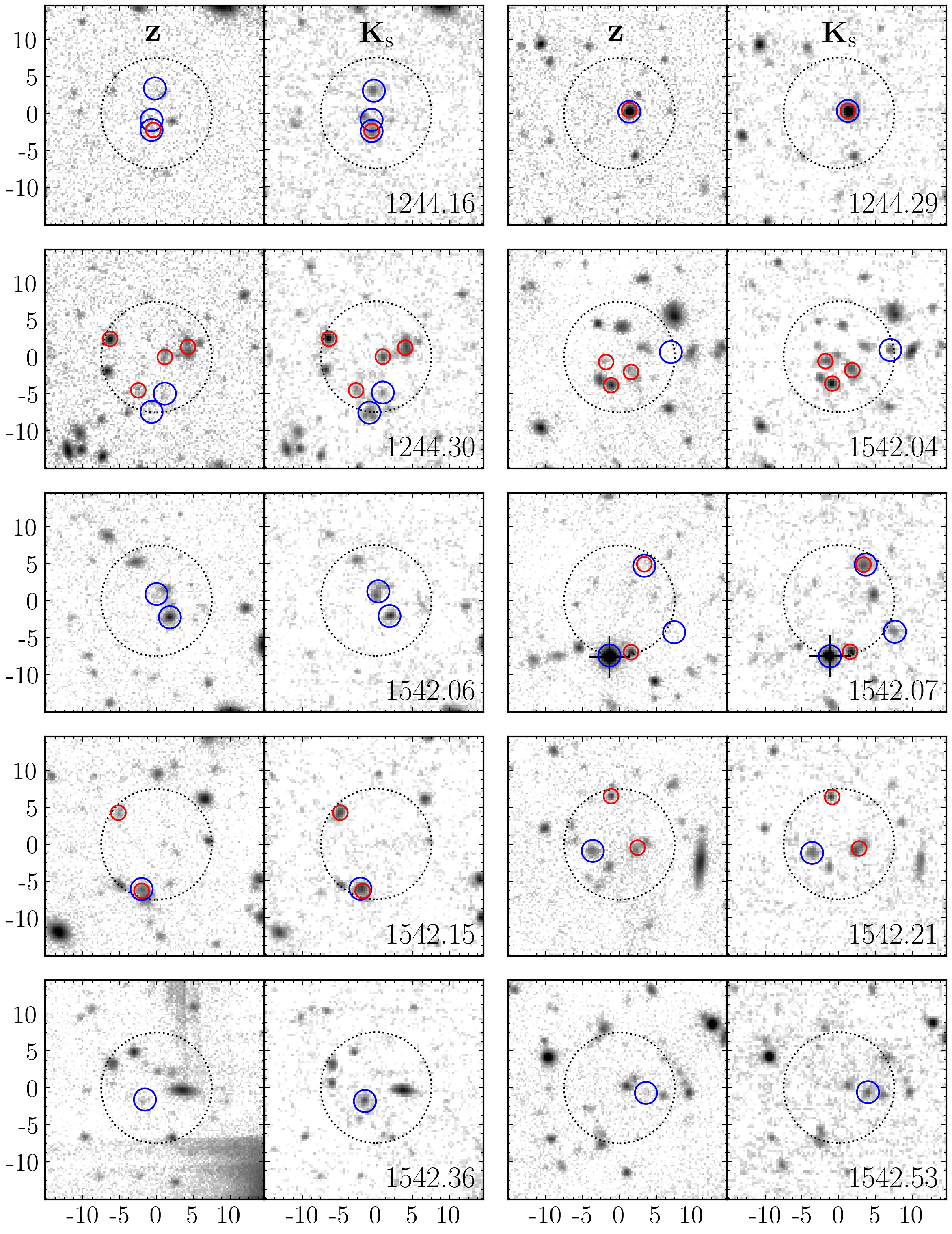}
    \caption{ The $z$ and $K_{\rm s}$-band stamp images of the ten HAE beam-matched SMGs in BOSS1244 and BOSS1542. There are totally fourteen HAEs are located within the SCUBA-2 850\,$\micron$ beam (dotted circles) of the ten SMGs. 
    Blue circles mark  the HAE candidates, and  red circles mark objects with $z-K_{\rm s}>1.5$. The ID numbers of the SMGs are given by their S/N in the catalogues.  For SMG 1542.07,  black crosses mark one HAE  at $z=2.24$ confirmed to be a quasar.}
    \label{fig:haes}
\end{figure*}

We examine the number density of the $z$ and $K_{\rm s}$-band sources as a function of distance from the central position of SMGs, and show the results in Figure~\ref{fig:radius}. For a comparison, we derive the average number density of the $z$ and $K_{\rm s}$-detected objects. This is done by randomly selecting 10\,000 positions in the $z$ and $K_{\rm s}$  images and calculating the number density as a function of radius centred at given positions.  Dashed lines show the reference results in Figure~\ref{fig:radius}. Note that the number density of the $z$-band sources is higher than that of $K_{\rm s}$-band sources as the $z$ images are deeper than the $K_{\rm s}$ images.  However, the source density at $R<2\arcsec$ is higher in $K_{\rm s}$ than in $z$, while at $R>2\arcsec$ it becomes the opposite.  It can be clearly seen that the number density of either $z$ or $K_{\rm s}$-band sources decreases from the central position of the SCUBA-2 sources to the outer radii, and close to the reference level at the radii of $>$10$\arcsec$.  It is obvious that the optical/NIR sources within $R<2\arcsec$ around the position of SMGs have the highest probability to be the true counterparts. Our results show that SMGs are less affected by dust in the $K_{\rm s}$ band and thus increase the chance to be detected, although this depends on the depths of the $z$ and $K_{\rm s}$ imaging. 

Interestingly, BOSS1244 exhibits a higher source density within $R<3\arcsec$ than BOSS1542. This difference can be seen in both $z$ and $K_{\rm s}$, suggesting that  SMGs in BOSS1244 are statistically brighter in the optical/NIR than SMGs in BOSS1542.  We attribute this to the excess at the bright end of the $z$ and $K_{\rm s}$ magnitude distributions in BOSS1244, as we can see a higher fraction of bright sources in these two bands in the upper panel of Figure~\ref{fig:color}.  
At $3\arcsec < R <12\arcsec$ there still a non-negligible excess of sources in $z$ and $K_{\rm s}$ relative to the reference level. We attribute this excess to a stronger clustering environment that SMGs live in.  The  excess of the number density  from the centre to  $R=13\arcsec$ corresponds $1.8\pm0.3$ ($1.4\pm0.3$) objects in $z$ down to $z=25.1$\,mag and $1.6\pm0.2$ ($1.3\pm0.2$) objects in $K_{\rm s}$  down to $K_{\rm s}=23.2$\,mag in BOSS1244 (BOSS1542). The source excess in $K_{\rm s}$  is smaller, compared with the estimate of 2.0 objects within $R=13\arcsec$ around the SCUBA-2 positions given in \cite{Simpson2019}. We caution that our results of the source excess has relatively large uncertainties due to the limited number of SMGs in our samples. 

We also cross match our samples of SMGs with HAE candidates. Three SMGs in BOSS1244 and seven SMGs in BOSS1542 are found to have the HAE candidates within the SCUBA-2 beam size.  Figure~\ref{fig:haes} demonstrates the $z$ and $K_{\rm s}$ images of the ten SMGs. HAEs and sources with $z-K_{\rm s}>1.5$ are marked. Three of these ten SMGs have more than one HAE candidates. Thus the fraction of multiplicity for SMGs detected with single-dish telescopes is thirty\,per\,cent when only considering the HAEs, slightly lower than the fraction (3 out of 7) around the Spiderweb galaxy \citep{Dannerbauer2014} and consistent with the estimates of multiplicity in the literature \citep{Hodge2013, Stach2018}.  
These groups of multiple HAEs are very likely merging galaxies, implying that a high fraction of SMGs  may be merging galaxies.   Among them, SMG 1244.16 matches a pair of HAEs located at the edge of the SW structure in BOSS1244. The two HAEs have a projected distance of $1\farcs2$.  In BOSS1542, there are  two HAEs and one quasar at $z=2.234$ form a group near the position of SMG 1542.07. SMG 1542.07 is found to match an HAE and a quasar in this group but separate $7\farcs7$ from the quasar (see black crosses in Figure~\ref{fig:density}). We plot all three objects around SMG 1542.07 in Figure~\ref{fig:haes} despite the separation of an HAE is slightly larger than half of the SCUBA-2 850\,$\micron$ beam size.  If confirmed to be the same object, the host of this quasar must be a dusty starburst galaxy. 

There are four HAE candidates located within the SCUBA-2 beam of four SMGs (SMG1244.16, SMG1244.29, SMG1542.07 and SMG1542.15) with $z-K_{\rm s}>1.5$,  being very likely to be the dusty objects. We adopt the method of corrected Poisson probability (i.e., p-value) to quantify the robustness of SMG counterpart identification, which was first done by \cite{Dannerbauer2014} for HAEs. 
The calculation of the p-values was reported by \cite{Downes1986} and described as follows:
\begin{equation}
    p = 1 - \exp (-\pi n \theta^2),
    \label{p-value}
\end{equation}
where $n$ represents the surface density of the HAE candidates within the effective area of our SCUBA-2 observations and $\theta$ is the separation between the HAE candidates and the submillimeter detections. We adopt the HAE surface density of 0.47 arcmin$^{-2}$ and 0.42 arcmin$^{-2}$ in BOSS1244 and BOSS1542 calculated by \cite{Zheng2021}. 
Totally, fourteen HAEs located within the beams of ten SCUBA-2 sources meet the criteria (p < 0.05) of the secure counterparts \citep{Biggs2011, Dannerbauer2014}, we suggest that these ten SMGs are well identified by HAEs in BOSS1244 and BOSS1542.  There are ten out of 97 SMGs are associated with the HAEs in BOSS1244 and BOSS1542,  lower than the fraction (seven out of eleven SMGs) reported in \cite{Dannerbauer2014}. 


\section{SUMMARY AND DISCUSSION} \label{sec:summary}

We present the results of our deep 850\,$\micron$ survey of two  spectroscopically-confirmed massive protoclusters of galaxies at $z=2.24$,  BOSS1244 and BOSS1542. The observations were carried out with JCMT/SCUBA-2. The final 850\,$\micron$  mosaic science maps reach a median sensitivity of $1.21 (1.12)$\,mJy\,beam$^{-1}$ within an effective area of 264\,arcmin$^2$ in BOSS1244 (BOSS1542). There are 43 and 54 submillimetre sources detected at the 4$\sigma$ level  in BOSS1244 and BOSS1542, with the deboosted flux range of  $S_{850}=2.8-$16.6\,mJy. We estimate the false-detection rate to be 7\,per\,cent at $>$4$\sigma$ and 2\,per\,cent at $>$4.5$\sigma$ from the recovery tests with the jackknife noise maps. Through the Monte Carlo simulations we demonstrate that the detection completeness reaches 50\,per\,cent (80\,per\,cent) at an intrinsic flux of 4.6\,mJy (6\,mJy) for BOSS1244 and at 4.1\,mJy (5.3\,mJy) for BOSS1542. The positional uncertainties of these SMGs are well confined within 3$\arcsec$ at $>$4$\sigma$.  These results  are consistent with the estimates from previous works on similar 850\,$\micron$ observations after accounting for the differences in depths and source density.

We construct the intrinsic number counts using our samples of 43 (54) SMGs detected in the BOSS1244 (BOSS1542) field. We find that the number density of SMGs in the two fields are $2.0\pm0.3$ and $2.1\pm0.2$ times the average of blank fields (S2CLS, S2COSMOS and NEPSC2).  
The overdensity  is even higher within the density peaks, with a factor of $5.1\pm0.9$ and $4.3\pm0.5$ in BOSS1244 and BOSS1542. It is reasonable to infer that nearly half of these SMGs belong to the $z=2.24$ overdensities while the rest are fore-/background SMGs.  Attributing the excess of SMGs to the  $z=2.24$ protoclusters,  we obtain an SMG volume density of $\sim(40$--$70) \times 10^{-5}$\,cMpc$^{-3}$,  being 15--30 times the average of the blank fields. The excess of SMGs with the overdensity of $2-$4 have been found in the Spiderweb protocluster and around  HzRGs  \citep{Dannerbauer2014, Zeballos2018}, as well as around WISE-selected AGNs and ELANs, with a factor of $2-$6 \citep{Jones2017, Nowotka2022}. Our results clearly show that BOSS1244 and BOSS1542 are extreme overdensities at $z>2$.  

Strikingly, the spatial distributions of our sample SMGs show obvious offsets from the high-density regions of HAEs in both BOSS1244 and BOSS1542. These SMGs are located mostly in the outskirts of the high-density regions mapped by HAEs. Within the effective area of our 850\,$\micron$ observations, the bulk of SMGs surround the SW density peak in BOSS1244, and in BOSS1542 SMGs are apparently clustered off the ridge peak of the giant filamentary structure. Meanwhile, inside the high-density regions, few SMGs are present. 
 Additionally, the comparison of the distributions between SMGs and other populations are also reported. \cite{Dannerbauer2014} showed that the spatial distribution of SMGs related to the Spiderweb galaxy is similar to the filamentary structure traced by HAEs, without the centre overlapping. \cite{Arrigoni2018} reported the alignment between the two brightest SMGs and the LAE density peak in MAMMOTH-1 nebula, probably indicating the the protocluster core. In SSA22, SMGs are found to have similar distribution to that of LAEs according to their two point angular cross correlation function \citep{Tamura2009, Umehata2014}. More studies on the distributions between SMGs, HAEs and LAEs are needed to panoramically reveal the evolution of protoclusters. 
Given that SMGs at $z=2.24$ in our samples are extreme starbursts, the overabundance of SMGs in the outskirts suggests an enhancement of gas supplies and/or an increasing frequency for starburst triggering events (e.g., galaxy mergers); and the deficiency of SMGs  inside the high-density regions indicates a disfavour for violent galaxy star formation, probably due to the cut-off of gas supplies in the collapsed regime of growing dense structures. 

We point out that the SW high-density region in BOSS1244 is a dominant component  forming a galaxy cluster core. The surrounding galaxies are expected to infall into the dense core structure through connected filaments. In the transition regime where the infalling galaxies and gas enter the collapsed regime gravitationally bounded by the core structure, the interacting processes (e.g., accretion shock) would probably accelerate gas cooling and boost gas accretion; and tidal disruptions could easily trigger starburst in gas-rich galaxies \citep{Narayanan2015, Rost2021}. Therefore more extreme starbursts are expected to take place in the outskirts of the high-density regions.  
Similar scenarios that the infalling galaxies with cold gas accretion are more active than that in the center of the protoclusters are reported in \cite{Dannerbauer2014}. By then the infalling starbursts are probably quenched by the lack of gas accretion and the energetic AGN feedback, and eventually form massive ellipticals which are seen in the local clusters \citep{Shimakawa2018b}. 
Such a picture is also applicable to BOSS1542, in which the bending filamentary structure appears likely to fold and collapse into a  relaxed structure; and the galaxies in the regions off the ridge region could be supplied with more gas from the shrinking large-scale environment. Again, the lack of SMGs inside the high-density ridge region of BOSS1542 supports our argument on the suppression of violent starbursts in the collapsed regime of growing dense structures.
Our finding that the overabundance of SMGs emerge in the outskirts of high-density regions in the two massive protoclusters of galaxies at $z=2.24$ provides a first direct probe for the impact of the assembly of large-scale structures on galaxy formation.

From deep $z$ and $K_{\rm s}$ imaging we identify 36 (46) optical/NIR objects with $z-K_{\rm s}>1.5$ located within the SCUBA-2 850\,$\micron$ beam coverage of our sample of 43 (54) SMGs in BOSS1244 (BOSS1542). These red objects are more likely to be the SMG counterparts than blue objects. No counterpart candidates of $z-K_{\rm s}>1.5$ down to $z\sim25.1$\,mag and $K_{\rm s}\sim 23.2$\,mag (5$\sigma$)  are found for some of SMGs in our samples ($\sim$16\,per\,cent).  Statistically, the locations of SMGs contain more objects  in $z$ and $K_{\rm s}$ relative to the places without SMGs.  The excesses  correspond to 1.8 (1.6) objects in $z$ and 1.4 (1.3) objects in $K_{\rm s}$ for BOSS1244 (BOSS1542). The cross matching between SMGs and HAE candidates  gives 6 (8) HAE candidates located around 3 (7) SMGs within the 850\,$\micron$ beam  in BOSS1244 (BOSS1542). Three of the ten SMGs meet more than one HAE candidate, suggesting a  multiplicity fraction ($\sim$30\,per\,cent) consistent with that given in literature \citep[e.g.][]{Stach2018}. In addition,  four of the ten SMGs have matched HAE candidates with $z-K_{\rm s}>1.5$,  indicating them very likely to be the actual counterparts.

We argue that more spectroscopic and imaging observations in the optical/NIR and (sub)millimetre bands will enable to obtain precise redshifts for different galaxy populations, map substructures and large-scale kinematics,  and characterize the detailed properties of member galaxies in these extremely massive protoclusters and surrounding cosmic web.  Doing so will draw a comprehensive picture for the formation of galaxies in connections with the assembly of large-scale structures.

\section*{Acknowledgements}

We are grateful to the anonymous referee in clarifying a number of important points and thus improving this manuscript.
This work is supported by the National Key R\&D Program of China (2017YFA0402703), the National Science Foundation of China (11773076 and 12073078),  the Major Science and Technology Project of Qinghai Province (2019-ZJ-A10), the science research grants from the China Manned Space Project with NO. CMS-CSST-2021-A02, CMS-CSST-2021-A04 and CMS-CSST-2021-A07, and the Chinese Academy of Sciences (CAS) through a China-Chile Joint Research Fund (CCJRF \#1809) administered by the CAS South America Centre for Astronomy (CASSACA). 
Y.G. acknowledges support by National Key Basic R\&D Program of China (2017YFA0402704) and NSFC grant 12033004.
HD acknowledgements financial support from the Agencia Estatal de Investigación del Ministerio de Ciencia e Innovación (AEI-MCINN) under grant (La evolución de los cíumulos de galaxias desde el amanecer hasta el mediodía cósmico) with reference (PID2019-105776GB-I00/DOI:10.13039/501100011033) and from the ACIISI, Consejería de Economía, Conocimiento y Empleo del Gobierno de Canarias and the European Regional Development Fund (ERDF) under grant with reference PROID2020010107.

The James Clerk Maxwell Telescope is operated by the East Asian Observatory on behalf of The National Astronomical Observatory of Japan; Academia Sinica Institute of Astronomy and Astrophysics; the Korea Astronomy and Space Science Institute; the National Astronomical Research Institute of Thailand; Center for Astronomical Mega-Science (as well as the National Key R\&D Program of China with No. 2017YFA0402700). Additional funding support is provided by the Science and Technology Facilities Council of the United Kingdom and participating universities and organizations in the United Kingdom and Canada. 
Additional funds for the construction of SCUBA-2 were provided by the Canada Foundation for Innovation.

\section*{Data Availability}

The data underlying this article will be shared on reasonable request
to the corresponding author.



{}

\label{lastpage}
\end{document}